\documentclass{aa}
\usepackage[varg]{txfonts}
\usepackage[utf8]{inputenc}
\usepackage{comment}
\usepackage{amsmath}
\usepackage{graphicx}
\usepackage{gensymb}
\usepackage[colorlinks=true,linkcolor=blue,citecolor=blue]{hyperref}
\usepackage{xspace}
\usepackage{geometry}  
\usepackage{lipsum}    
\usepackage{capt-of}

\usepackage{gensymb}
\DeclareRobustCommand{\ion}[2]{\textup{#1\,\textsc{\lowercase{#2}}}}
\newcommand{\jwst}{\emph{JWST}\xspace}

\newcommand{\sirocco}{\textsc{Sirocco}\xspace}

\usepackage{orcidlink}

\usepackage{natbib}
\bibpunct{(}{)}{;}{a}{}{,} 

\begin{document}

\title{UV spectra of little red dots: \\Broad-line N/C, gas recycling and black hole masses}

\authorrunning{Sneppen et al.}

\author{Albert Sneppen\inst{\ref{addr:IoA},\ref{addr:kavli},\ref{addr:DAWN},\ref{addr:jagtvej}}\thanks{Corresponding author: \email{as3980@cam.ac.uk}}\orcidlink{0000-0002-5460-6126}, Darach Watson\inst{\ref{addr:DAWN},\ref{addr:jagtvej}}\orcidlink{0000-0002-4465-8264}, Lucy R. Ivey\inst{\ref{addr:kavli},\ref{addr:Cav}}\orcidlink{0009-0002-5105-1222}, Kasper E. Heintz\inst{\ref{addr:DAWN},\ref{addr:dtu}}\orcidlink{0000-0002-9389-7413}, Alex J. Cameron\inst{\ref{addr:DAWN},\ref{addr:jagtvej}}\orcidlink{0000-0002-0450-7306}, \\Joris Witstok\inst{\ref{addr:DAWN},\ref{addr:jagtvej}}\orcidlink{0000-0002-7595-121X}, Vadim Rusakov \inst{\ref{addr:man}}, James H. Matthews\inst{\ref{addr:oxford}}\orcidlink{0000-0002-3493-7737}, Rasmus Damgaard\inst{\ref{addr:DAWN},\ref{addr:jagtvej}}, Georgios Nikopoulos\inst{\ref{addr:DAWN},\ref{addr:jagtvej}}\orcidlink{0009-0004-6791-9246} and Vasily Belokurov \inst{\ref{addr:IoA}}
}

\institute{
Institute of Astronomy, University of Cambridge, Madingley Road, Cambridge, CB3 0HA, UK \label{addr:IoA}
\and
Kavli Institute for Cosmology, University of Cambridge, Madingley Road, Cambridge, CB3 0HA, UK \label{addr:kavli}
\and
Cosmic Dawn Center (DAWN)\label{addr:DAWN}
\and
Niels Bohr Institute, University of Copenhagen, Jagtvej 128, DK-2200, Copenhagen N, Denmark \label{addr:jagtvej} 
\and
Cavendish Laboratory, University of Cambridge, 19 JJ Thomson Avenue, Cambridge CB3 0HE, UK \label{addr:Cav}
\and
DTU Space, Technical University of Denmark, Elektrovej 327, DK2800 Kgs. Lyngby, Denmark \label{addr:dtu}
\and
Jodrell Bank Centre for Astrophysics, University of Manchester, Oxford Road, Manchester M13 9PL, UK\label{addr:man}
\and 
Astrophysics, Department of Physics, University of Oxford, Keble Road, 
Oxford OX1 3RH, UK\label{addr:oxford}
}

\date{Received date /
Accepted date }

\abstract{ 
    Radiative-transfer models of supermassive black holes accreting inside dense gas cocoons reproduce the X-ray-to-near-infrared spectral energy distribution of `little red dots' (LRDs) using the code \sirocco. In the ultraviolet, these models also predict weak broad emission lines and a series of metal absorption lines, which deep spectroscopy of the \jwst SPURS survey now detects in several LRDs. Here we use the models to interpret these features. We find  that N and C ions of equal charge (3+ or 2+) occupy largely the same gas in the simulations, such that the combination of the broad \ion{N}{iv}]/\ion{C}{iv} and \ion{N}{iii}]/\ion{C}{iii}] flux ratios simultaneously constrains both the electron density and the N/C abundance ratio of the gas in the LRD proper. The nitrogen and carbon emission of SPURS-GN-2, an LRD at $z=7.19$, requires an order of magnitude nitrogen excess compared to the solar value, ${\rm N/C = 12_{-2}^{+3} \,(N/C)_\odot}$, in the broad-line gas with electron density \(\log{n_e} = 9.7^{+{0.3}}_{-0.5}\)\,cm\(^{-3}\). Thus, around half the broad line CNO nuclei have passed through hydrogen burning. Because the ionised gas is replaced on a century timescale, sustaining this enrichment favours gas recycling. UV absorption lines test this independently. In two of the four LRDs, line blueshift velocities clearly increase with the photon energy required to ionise the absorber gas. In the models the absorption lines are spatially stratified, indicating that the gas is launched near the escape speed and slowed by gravity. In GN-2, where the measurement can be made, the inferred deceleration requires radiation pressure close to the Eddington limit of a $10^{6-7}\,\mathrm{M_\odot}$ black hole. The lines are also typically strongly saturated, making inferences of direct abundance ratios misleading.
    }

\maketitle

\section{Introduction}

A candidate population of supermassive black holes embedded in dense gas, the so-called `Little Red Dots' \citep[LRDs;][]{Matthee2024,Harikane2023,Kocevski2023}, is characterised, among other properties, by their `V-shaped' spectra with blue ultraviolet (UV) and red optical colours. The inflection near the Balmer break wavelength \citep{Setton2024}, in some cases more prominent than possible with stellar populations \citep{deGraaff2025,Naidu2025}, is likely due to UV absorption by enshrouding gas \citep{Inayoshi2024,Rusakov2025,Sneppen2026,Matthee2026}. A population of `little blue dots' (LBDs) has also been discovered with properties similar to LRDs \citep[e.g.][]{Brazzini2026,Rusakov2025,Geris2026,Matthee2026}, and has been interpreted as LRDs with lower column densities \citep{Sneppen2026,Sneppen2026d,Mascia2026}. We therefore include these objects under the general term LRD unless otherwise specified. 

The UV light from LRDs is sometimes extended, indicating a significant host galaxy contribution at these wavelengths \citep[e.g.][]{Rinaldi2024,Cloonan2026,Durodola2026,Sun2026,Ishikawa2026}. Several properties, however, require that some UV emission from the LRD phenomenon itself must escape: LRD UV emission is typically redder than star-forming galaxies at the same redshifts and the observed damped Ly$\alpha$ downturn is correlated with the strength of the Balmer break \citep{Ando2026} -- the UV light is therefore tied to the nuclear gas structure that produces the break, indicating an origin near the gas-enshrouded nucleus rather than in the host; furthermore, while the population as a whole seems unusually deficient in classical high-ionisation AGN UV lines \citep{Lambrides2024,Tang2025,wang2025missing}, moderately high-ionisation lines have nevertheless been detected in deep spectra, e.g.\ \ion{N}{IV}, \ion{C}{IV} and \ion{He}{II}, demonstrating that photons with energies above \(\sim\!50\)\,eV are present in at least some LRDs \citep{Akins2025b,Tang2025,Brazzini2026,Tang2026b}; at a population level, the Ly$\alpha$ luminosity is comparable to similar redshift star-forming galaxies \citep{Ji2026b}, but broad Ly$\alpha$ lines have been detected in a small number of individual objects \citep{Ji2026,Tang2026,Kageura2026,Morishita2026}. 

\begin{figure*}
\vspace{-0.4cm}
\begin{center}
    \includegraphics[angle=0,width=\textwidth]{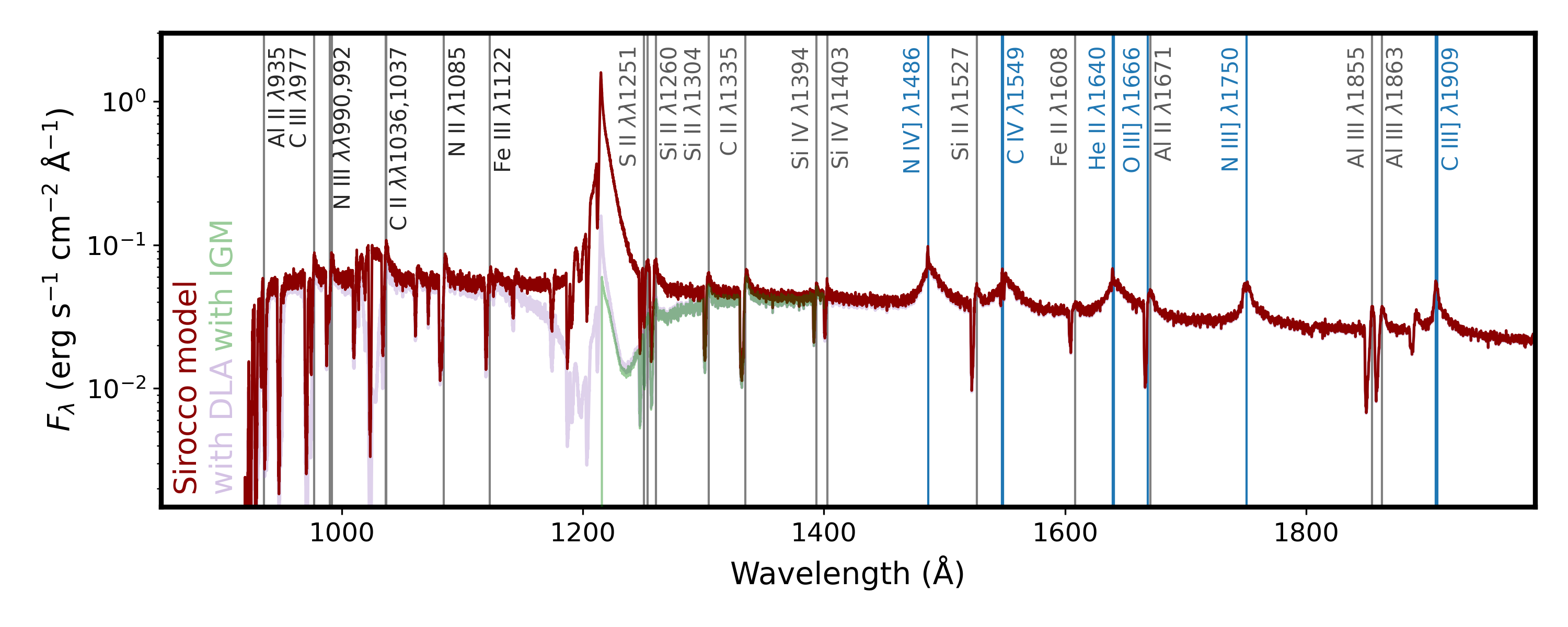}
\end{center}
\vspace{-0.8cm}
\caption{Rest-UV \sirocco model spectrum \citep[adapted from Fig. 1 in][]{Sneppen2026d} with the largest emission (blue) and absorption (grey) features annotated. The Lyman series is not annotated. By post-processing the model we show the impact of a damped Ly$\alpha$ wing $N_{\rm \ion{H}{I}}=10^{22.5} {\rm cm}^{-2}$ and large covering-factor $f_{\rm cov}=0.9$ in purple, as well as IGM absorption at $z=7$ and $X_{\ion{H}{I}}=0.5$ in green. The sole update to the model compared to \citet{Sneppen2026d} is an increased N/C abundance. 
}
\label{fig:UV}
\end{figure*}

Self-consistent radiative transfer modelling with the \sirocco code has demonstrated that AGN emission mediated through dense gas flows around a supermassive BH could reproduce the dearth of X-ray emission, the UV through NIR SED, the ensemble of emission and absorption lines, as well as unify the diverse populations observed \citep{Sneppen2026,Sneppen2026c}. In the UV, such modelling has predicted weak broad emission lines \citep{Sneppen2026d,Mascia2026,Kaaz2026} and rich absorption lines from metals in the enshrouding gas (see Fig.~\ref{fig:UV}). Now deep rest-UV grating spectroscopy from the \jwst SPURS program of four LRDs at \(z=6.7{-}7.2\) and $M_{UV}<{-}20$ \citep[GO-9214, PIs Mason \& Stark,][]{Chen2026spurs} has provided stringent constraints on metal absorption features \citep{Tang2026b,Kokorev2026}. All four LRDs, labelled GN-2, GN-29, GN-2004 in GOODS-N and CEERS-7902 in EGS, show nitrogen-enhanced narrow-line gas, are UV-bright $M_\mathrm{UV} = -20.4$ to $-21.1$ and span the blue end of the LRD colour distribution \citep[$\beta_\mathrm{opt} = 0.06{-}0.48$][]{Tang2026b}. Notably, such nitrogen-rich gas, a hallmark of a growing class of high-redshift N-emitters \citep{Cameron2023,Schaerer2024,Topping2024}, links the dense nuclear environments of LRDs to the sites of hydrogen-burnt material \citep[e.g.\ see discussion in][]{Torralba2025_cola}.

In this paper, we revisit these \sirocco models to expand on the detailed constraints permitted by the absorbers and the emission features. In the context of the SPURS LRDs, this allows us to i) measure the nitrogen overabundance of the broad-line gas proper and ii) probe the radial velocity law of the gas flow. 


\section{Methods}
We employ \sirocco \citep{long2002,Matthews2025} to model LRDs with self-consistent radiative transfer and adopt the axisymmetric inflow/outflow gas-cocooned AGN models described in previous papers (i.e., \citet{Sneppen2026b,Sneppen2026d,Sneppen2026}, see also \citet{Mascia2026,Kaaz2026}) and detailed in App.~\ref{app:bh}. These papers employ a standard \sirocco line list,
where for instance the resonance doublet \ion{C}{iv}\;$\lambda\lambda$1548,1551, like all metal lines, is treated as a collisionally excited two-level atom in the Sobolev approximation with resonance scattering and self-absorption redistributing photons. This line list contains no UV semi-forbidden lines other than \ion{C}{iii}]~$\lambda\lambda$1907,1909. We therefore expanded the line-list (e.g.\ \ion{N}{ii}]~$\lambda2143$, \ion{N}{iii}]~$\lambda1750$, \ion{N}{iv}]~$\lambda1486$, \ion{O}{iii}]~$\lambda\lambda$1661,1666, \ion{O}{iv}]~$\lambda1401$, \ion{O}{v}]~$\lambda1218$, \ion{C}{ii}]~$\lambda2326$) using `simple-atom' transitions with radiative rates and collision strengths from \textsc{chianti} v10.1 \citep{Dere1997,DelZanna2021}. This treatment captures their emissivity and its collisional quenching. We assume a solar relative abundance pattern with $Z=0.1 \,Z_\odot$ \citep{Nikopoulos2026}, but explore N/C variations, motivated by growing evidence for nitrogen enhancement in at least some LRDs \citep{Labbe2024,Morishita2026,Papovich2026,Tang2026b,Torralba2025_cola}. 

For each relevant ion and line in \sirocco spectra, we fit Gaussian absorption lines to determine the line centroids as done in \citet{Tang2026b}. 
The cocoons in our models are observed to be ionisation stratified, which means that the velocities of the absorption lines of the different metal ions effectively probe the radial kinematics of the gas. We adopt a ballistic radial velocity-law, $v_R(r)^{2} = v_R(\infty)^{2} + 2GM_{\rm eff}/r$, characterised by a wind driven near the escape velocity (i.e.\ $v(r)\gg v({\infty})$) of an effective potential $GM_{\rm eff}/r$, reminiscent of super-Eddington winds proposed for LRDs \citep{Kaaz2026}. We favour this model as it is i) theoretically motivated by the dynamics of such gas and ii) explains the observed ordering of the absorbing ions in some LRDs. The observed absorption ordering constrains the ionisation state as a function of velocity not radius, so an accelerating wind with ionisation parameter rising outward would produce a similar trend. However, low-ionisation troughs and Balmer breaks imposed on the spectrum require neutral gas outside the continuum-forming region, favouring the solution where ionisation state declines as a function of radius outwards.

\begin{figure*}[!t]
\vspace{-0.4cm}
\begin{center}
    \includegraphics[angle=0,width=\textwidth]{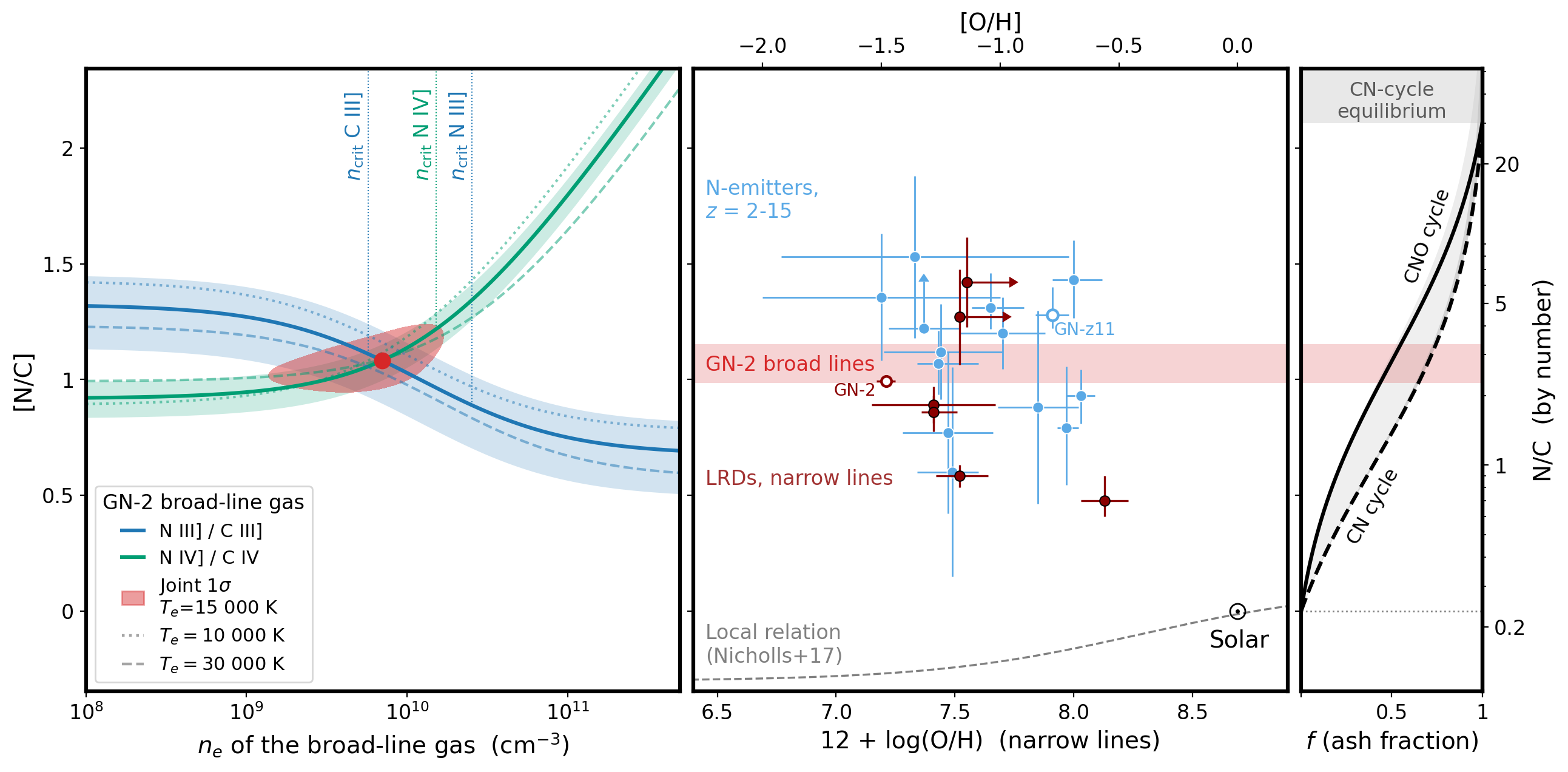}
\end{center}
\vspace{-0.5cm}
\caption{
Nitrogen abundance in GN-2 broad-line gas. 
Left panel: N/C versus electron-density from the \ion{N}{iii}]/\ion{C}{iii}] (blue) and \ion{N}{iv}]/\ion{C}{iv} (green) flux ratios \citep[see Fig.~\ref{fig:gn2fits} and][]{Tang2026b} and atomic data alone. Red marks the joint constraint, vertical lines show the critical densities and dotted and dashed curves show the modest temperature dependence (from 10\,000\,K-30\,000\,K with fiducial $T_e=15\,000$\,K). Middle panel: N/C-abundance in broad-line gas (red bar), in narrow-line LRDs \citep[red,][]{Ji2026_N,Papovich2026,Tang2026b} and in high-redshift N-emitters \citep[blue, compilation from][]{Schaerer2024,Naidu2026}. Open circles highlight GN-2 and GN-z11 and the dashed line shows the locus of low-$z$ star-forming galaxies \citep{Nicholls2017}.
Right panel: N/C of solar-ratio gas mixed with a fraction $f$ of hydrogen-burning ash for the CN cycle (dashed) or the CNO cycle (solid), where the GN-2 broad-line ratios require $f\gtrsim0.5$.
}\label{fig:nc_density}
\end{figure*}

Here we use the self-consistently solved ionisation structure in \sirocco to extract the ${\rm N_{\rm H\,I}}$ and gauge the cocoon's own potential imprint on the damping wing. 
Specifically, we add the wing in post-processing where the column is integrated along the radial ray nearest the sightline, $N_{\ion{H}{i}}=\sum x_{\ion{H}{i}}\,n_{\rm H}\,f\,\Delta r$. The optical depth is a sum of Voigt profiles for Ly$\alpha$--Ly10 \citep{Morton2003}, and is applied with partial covering, $F = F_{\rm model}\left[\, f_{\rm cov}\, e^{-\tau} + (1-f_{\rm cov}) \,\right].$ The treatment of the neutral column as a slab is an approximation motivated by an interpretation of the damped Ly$\alpha$ agnostic to whether the damping wing arises from the cocoon reservoir itself or from a foreground (galactic) screen. 
The hydrogen two-photon continuum, which can mimic a downturn blueward of $1400$\,\AA\ \citep{Cameron2024}, is not computed by \sirocco's $n$-resolved hydrogen atom. Regardless, the observed spectra show a Ly$\alpha$ trough rather than the gradual two-photon turnover \citep{Tang2026b}. We note that the Sobolev approximation used in \sirocco makes the emergent Ly$\alpha$ line-shape potentially unreliable, which requires dedicated Ly$\alpha$ Monte Carlo treatment, although it is still possible that electron-scattering dominates line-broadening in the ionised medium for very low neutral fractions. 

A key assumption throughout this \sirocco modelling is that the gas-enshrouded AGN produces the observed UV-features of absorption and broad emission. For the interpretation of differential absorber kinematics, it further assumes the black hole dominates the gravitational potential within the absorbing region. Alternate frameworks, such as offset between high and low ions for star-forming galaxies with multiphase winds \citep{Vasan2026,Li2026} or broad N\,\textsc{iv}] potentially from the winds of very massive stars \citep{Chen2026}, are outside the purview of these models (for a detailed discussion of a stellar UV see \citet{Tang2026b}). We emphasise that the ensemble of UV absorption and emission probes discussed here was predicted for the gas-enshrouded AGN phase from models constrained by the optical spectra \citep{Sneppen2026,Sneppen2026d}, not features fitted to the SPURS data.


\section{Results}
Fig.~\ref{fig:UV} shows the spectrum of a model published previously \citep{Sneppen2026d}, which reproduces typical optical features of LRD-like objects (e.g.\ Balmer broad line-width and line-absorption). Here we focus on the UV part, but generally emphasise that the modest metallicity at $Z=0.1\,Z_\odot$ is sufficient to produce both the observed metal features and hydrogen lines. Next we will address in turn the model predictions and the observational comparison of i) broad metal emission and ii) narrow metal absorption.

\subsection{Broad emission features -- N/C overabundance} 
Several broad UV emission features of similar width, modest intensity and at times with significant self-absorption are produced in the models \citep{Sneppen2026d}. By including electron and resonant scattering effects \citep[e.g.\ as likely relevant for \ion{C}{IV};][]{Lim2026,Li2026a}, \sirocco allows us to infer constraints on elemental abundance ratios of the broad-line emitting gas. \citet{Tang2026b} reported broad components of \ion{N}{IV}]\,$\lambda1486$, \ion{C}{IV}\,$\lambda1549$, \ion{N}{III]}\,$\lambda1750$ and \ion{C}{III]}\,$\lambda1909$ in GN‑2, while our independent fits agree in flux ratios within $1\sigma$ and additionally give a $2\sigma$ upper limit on broad \ion{O}{III]}\,$\lambda1666$ (see App.~\ref{app:abund}).

In GN-2, the broad flux ratio \ion{N}{IV}\,$\lambda1486$/\ion{C}{IV}\,$\lambda1549$ is around eight times the expectation from solar N/C (at $T_{\rm e}\!=\!15\,000$\,K, low density $n_e\!<\!10^{9} \,{\rm cm^{-3}}$, see App.~\ref{app:abund}). This ratio measures N/C directly, because both ions are created at $\approx47.5$\,eV and in our \sirocco models occupy similar gas with emission-weighted ionic fractions $x({\rm \ion{N}{IV}})/x({\rm \ion{C}{IV}})=0.8{-}1.0$\footnote{With ${x({\rm \ion{N}{iv}})=n(\ion{N}{iv})/n({\rm N})}$, ${x({\rm \ion{C}{iv}})=n(\ion{C}{iv})/n({\rm C})}$, etc.}. As \ion{C}{IV} is a permitted line, only the nitrogen line is collisionally quenched with a critical density $n_{\rm crit}$(\ion{N}{IV]})=$10^{10.2}\,{\rm cm^{-3}}$. The N/C is thus unambiguously super-solar and can only rise with density (${\rm N/C > 6\,(N/C)_{\odot}}$ at $2\sigma$ for any $n_e$, see Fig.~\ref{fig:nc_density} left panel). 

The broad \ion{N}{III]}\,$\lambda1750$/\ion{C}{III]}\,$\lambda1909$ ratio provides a complementary N/C at about 15--30 times the solar expectation (at $T_{\rm e}=15\,000$\,K and $n_e\!<\!10^9\,{\rm cm^{-3}}$). This is a lower limit with $x({\rm \ion{N}{III}})/x({\rm \ion{C}{III}})\!\lesssim\!1$ as the \ion{N}{III} zone (29.6--47.5\,eV) is enclosed within the \ion{C}{III} zone (24.4--47.9\,eV). This ratio moves in the opposite direction with increasing $n_e$, as the inferred N/C first decreases due to the critical density of the denominator, $n_{\rm crit}$(\ion{C}{III]})=$10^{9.8}\,{\rm cm}^{-3}$, and then saturates when both lines are quenched beyond $n_{\rm crit}$(\ion{N}{III]})=$10^{10.5}\,{\rm cm}^{-3}$. The two line-ratios are reconciled only if ${\rm N/C = 12_{-2}^{+3}\,(N/C)_\odot}$ and $n_e\!\sim\!10^{9.3}{-}10^{10}\,{\rm cm}^{-3}$ from atomic data arguments alone (see Fig.~\ref{fig:nc_density}, left panel). Although each ionisation pair is co-spatial, the \ion{N}{iv]} zone is denser than the \ion{N}{iii]} zone in \sirocco models. If both share one N/C, this ordering can only raise the inferred N/C up to $20\,({\rm N/C})_\odot$ and keeps the \ion{N}{iv]} zone at $n_e\sim10^{9.8}{-}10^{10.3}\,{\rm cm^{-3}}$. 

The density is bounded from above at $n_e\lesssim2\times10^{10}\,{\rm cm^{-3}}$ due to the broad-line \ion{N}{IV]} detection. This broad-line $n_e$ is consistent with the lower $n_e$-limit of the `narrow-line' gas \citep[$>10^{6}\,{\rm cm^{-3}}$ at $3\sigma$ from $\ion{C}{III}{]}/{[}\ion{C}{III}{]}$,][]{Tang2026b}, which could be the unscattered component of the same broad-line gas. The constraint is also consistent with our fiducial \sirocco model density, but does not permit very strong clumping in the spatial scales  modelled. In particular, given i) the upper limit on $n_e$, ii) the required columns for LRDs \citep[e.g.\ $N_e\!\sim\!10^{25}\,{\rm cm^{-2}}$ in][]{Sneppen2026d} and iii) the implied cocoon spatial scales, the volume filling-factor is $\gtrsim0.05\,(N_e/10^{25}\,{\rm cm^{-2}}) (R/10^{16} \,{\rm cm})^{-1}$. This clumping is within our space of models with clumping factors up to $\sim\!100$ explored in \citet{Sneppen2026}, but disfavours the most extreme clumped scenarios. 

Broad-line gas N/O provides a complementary albeit weaker constraint, but is nonetheless interesting as the intermediate processing from hydrogen burning can either keep its oxygen (the CN cycle) or have burnt it to nitrogen as well (the full CNO cycle). In the \sirocco models, a broad oxygen line O\,{\sc iii}]$\lambda1666$ is produced when there is both i) solar N/O and ii) a large \ion{N}{iv}] flux as observed. \ion{O}{iii} (35.1--54.9 eV) overlaps mostly with the gas of \ion{N}{iv}, with the majority of O\,{\sc iii}] emission coming from the denser N\,{\sc iv} zone of the \sirocco models, while the \ion{O}{iii} level itself is less density-quenched, with a critical density $n_{\rm crit}$(O\,{\sc iii}])\,$=\!10^{10.7}\,{\rm cm^{-3}}$, higher than the nitrogen lines. The observed broad N\,{\sc iv}] and N\,{\sc iii}] lines and the undetected broad O\,{\sc iii}]$\lambda1666$ (with $<6\times10^{-19}$ erg\,s$^{-1}$\,cm$^{-2}$ at $2\sigma$, see Fig.~\ref{fig:gn2fits}), together imply N/O\,$\gtrsim\!0.7$ (N/O$\,\gtrsim\;5$(N/O)$_\odot$), implying highly super-solar N/O as well as highly super-solar N/C.  

The GN-2 cocoon is thus likely as nitrogen overabundant as the narrow-line gas \citep{Tang2026b}, as well as nitrogen-loud quasars \citep{Baldwin2003,Matsuoka2017}, and the population of high-redshift N-emitters \citep[e.g.][]{Cameron2023,Schaerer2024,Topping2024,Topping2025,Naidu2026,Cameron2026,Rusakov2026b}, see Fig.~\ref{fig:nc_density}, middle panel. It is unlikely to be a coincidence of a small sample that all four SPURS LRDs show nitrogen emission \citep[measured in][]{Tang2026b}, whereas only 3 of 38 high-redshift galaxies with comparably deep rest-UV spectra show \ion{N}{iv]} \citep{Topping2025}, which corresponds to a chance probability of $3\times10^{-4}$ or $3.4\sigma$ significance. Indeed, \citet{Tang2026b} detect \ion{N}{iv}] or \ion{N}{iii}] in 12 of 18 LRDs with sufficiently deep archival spectra.
Such nitrogen-rich conditions, now determined to be present specifically in the broad-line gas for the first time, make a direct link between the LRD engines, formation-channel or environments \citep{Cameron2023,Chisholm2026,Tang2026b,Kokorev2026} and the nitrogen producers (e.g.\ Wolf-Rayet stars \citep{Kobayashi2024}, very massive stars \citep[VMS,][]{Crowther2010,Vink2023} or as of yet unobserved supermassive stars \citep[SMS,][]{Denissenkov2014,Charbonnel2023}). Tidal disruption events, a proposed LRD mechanism \citep{Bellovary2025} and a known nitrogen-rich source \citep[e.g.,][]{Yang2017}, are not considered here because this source would require i) regular ($\sim$ yearly) disruptions to supply the cocoon and ii) tidal forces act only on scales several orders of magnitude below the cocoon. These constraints suggest that the dense gas shares or possibly even produces, the nitrogen-rich reservoir of the region producing the narrow lines. It is worth emphasising again that the narrow lines where the N-excess is measured are fairly high-ionisation and appear to come from very high-density regions \citep[$\gtrsim10^{6}\,{\rm cm^{-3}}$, see Fig. 8 in][]{Tang2026b}, despite the fact that they are relatively narrow, so they are quite likely not to come from the host galaxy generally, but rather from the LRD cocoon or possible from some special region(s) in the host. It remains to be shown whether the lower density, lower ionisation regions of the bulk of the host galaxy also show the same N-excess. We emphasise that in general, such broad high-ionisation components are predicted by \sirocco, for the lowest-column density objects, where the least gas reprocessing leaves the bluest LRD \citep{Sneppen2026,Sneppen2026d,Matthee2026,Mascia2026}. The bluest SPURS LRD, GN-2, is exactly the LRD subclass where prominent UV broad lines are expected in \sirocco models.

\subsubsection*{The CNO-processed fraction, the nitrogen budget, and the requirement to recycle the gas}

The observed ${\rm N/C=12_{-2}^{+3} \,(N/C)_\odot}$ lies far above ordinary enrichment \citep{Nicholls2017} and far below the hydrogen burning equilibrium \citep[$\,\gtrsim\!100\,{\rm (N/C)_\odot}$, see][]{Arnould1999}, so this requires a mixture of processed and unprocessed material. The carbon share probes the processed fraction directly. CNO burning converts the polluters' own initial carbon and oxygen into nitrogen while conserving C+N+O \citep{Charbonnel2023,Higgins2025}. Fig.~\ref{fig:nc_density} (right panel), which mixes solar-scaled gas with ash at hydrogen-burning equilibrium, illustrates that the observed ${\rm N/C}$ implies $f\approx50\%$ has passed through the CNO cycle, or $f\approx70\%$ if only the CN cycle is considered. The absence of broad \ion{O}{iii}]$\lambda1666$ and the high implied N/O ratio favour the former (App.~\ref{app:abund}). In every other respect $f$ is a lower limit. Wolf-Rayet winds can add helium-burning carbon \citep{Charbonnel2023}, which would only lower N/C at fixed $f$. Metal-poor unprocessed gas starts with a lower N/C than the solar-scaled ratios adopted here, so the same observed ratio would require more processing. Ash that has not reached equilibrium would require more still. We emphasise these abundance ratios are largely independent of metallicity. From these extreme processed fractions, the limited UV luminosity, and the rapid turnover rate of the cocoon gas, we can infer whether the cocoon gas must be recycled, rather than purely fed-in and ejected, as follows. 

\begin{figure}[b]
\vspace{-0.15cm}
\begin{center}
    \includegraphics[angle=0,width=0.45\textwidth]{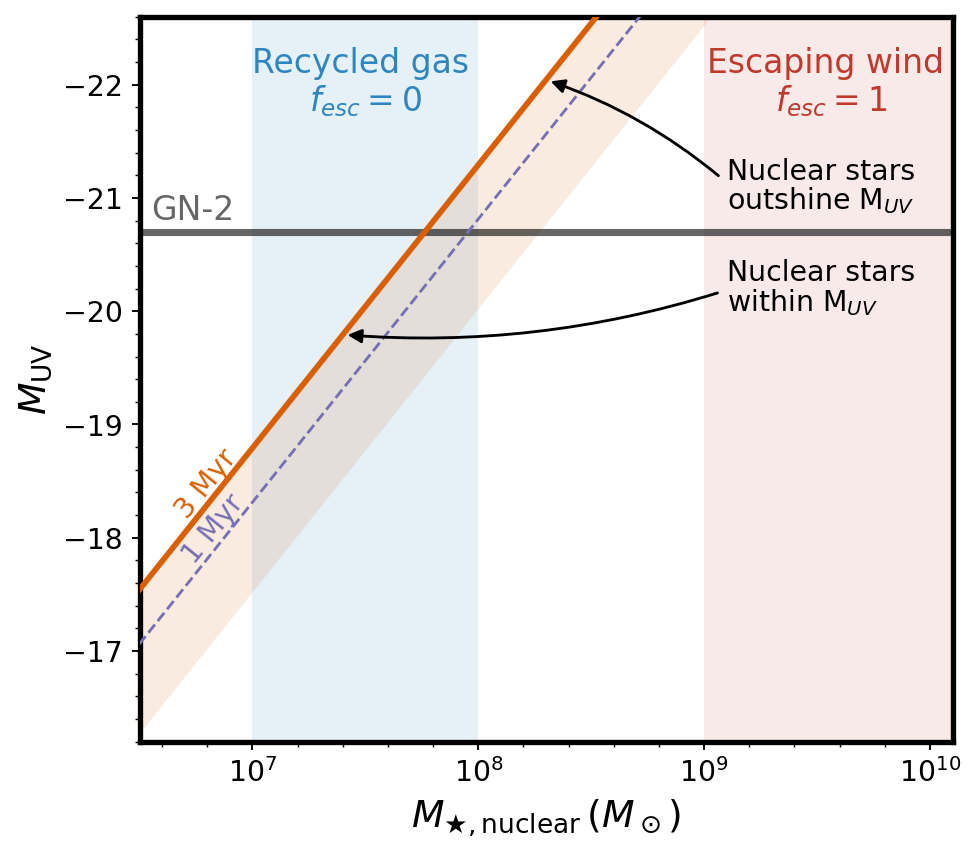}
\end{center}
\vspace{-0.55cm}
\caption{UV luminosity versus stellar mass for single stellar populations of 1 and 3\,Myr from \textsc{Bagpipes} \citep{Carnall2018} with a \citet{Kroupa2001} IMF \citep[shading shows dust reddening $0<A_V<0.5$,][]{Calzetti2000}. A continuous star formation over 10 Myr gives the same $M_{\rm UV}$ per unit mass as the 1 Myr burst. A stellar population older than this would dominate the abundances with supernova ejecta. The stars cannot be brighter than GN-2's $M_{\rm UV}$ \citep[horizontal line,][]{Tang2026b}. Vertical bands illustrate the nuclear stellar masses motivated by the nitrogen budget over $t_N=10\,$Myr for recycled gas (blue, Eq.~\ref{eq:m2} with $f_{\rm esc}=0$) and an escaping wind (red, $f_{\rm esc}=1$).} 
\label{fig:muv}
\end{figure}

\begin{figure}
\vspace{-0.05cm}
\begin{center}
    \includegraphics[angle=0,width=0.5\textwidth]{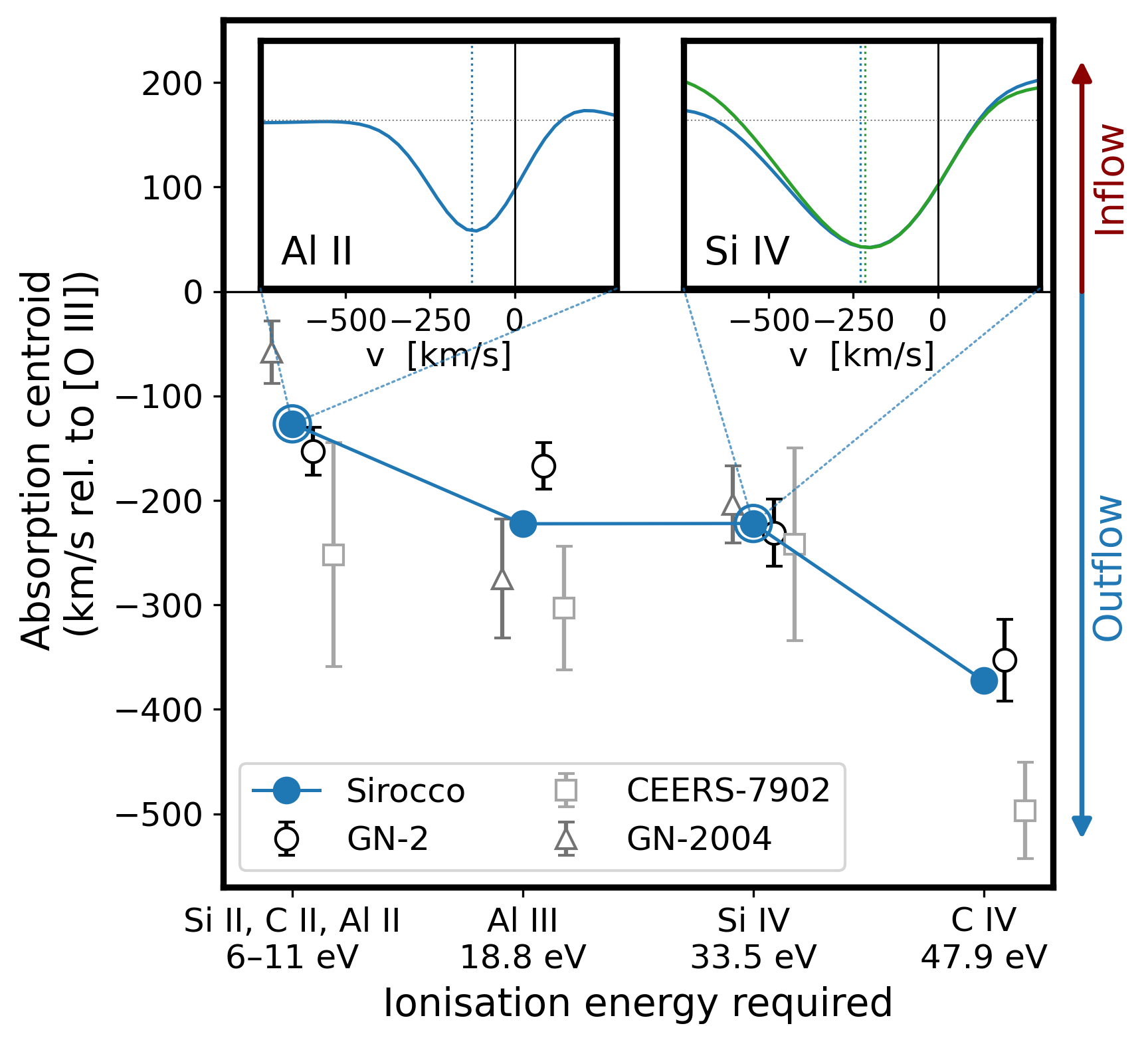}
\end{center}
\vspace{-0.55cm}
\caption{
UV absorption velocity centroids for ions of increasing ionisation potential. From \cite{Tang2026b}, we show the absorption line velocity centroids for GN-2, CEERS-7902 and GN-2004. The insets illustrate the absorption-line profiles from  \sirocco for the isolated lines of \ion{Al}{II} (at 1670.8\,\AA) and \ion{Si}{IV} (at 1393.8\,\AA\, and 1402.8\,\AA), where a decreasing radial velocity profile produces faster outflows for the more centrally located higher ionisation states.
}
\label{fig:abs_centroid}
\end{figure}

\begin{figure}
\vspace{-0.05cm}
\begin{center}
    \includegraphics[angle=0,width=0.45\textwidth]{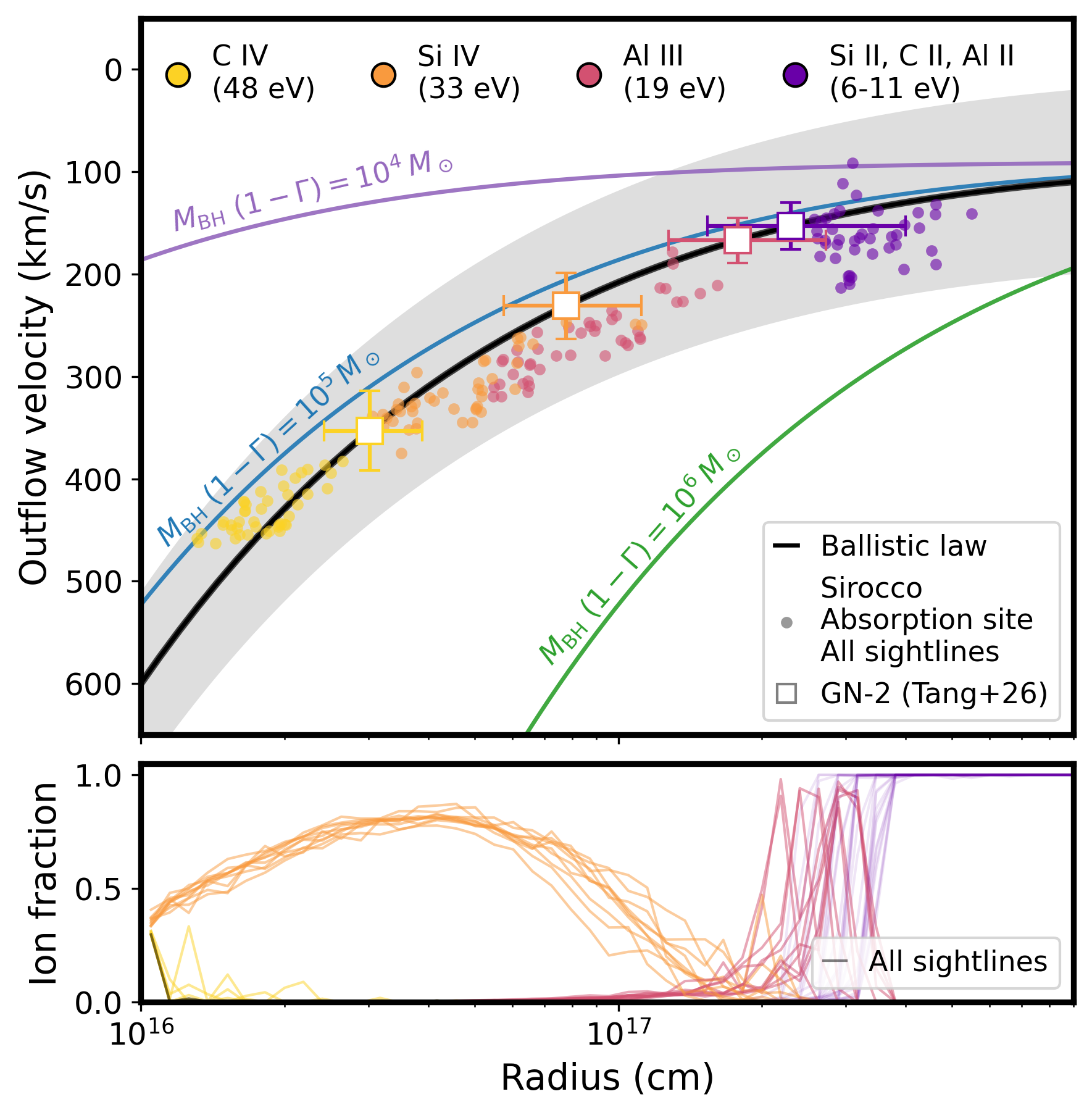}
\end{center}
\vspace{-0.65cm}
\caption{
Outflow velocity for species with different ionisation potentials as a function of radial distance. Top panel: The \sirocco model prediction for a radially-slowing velocity field is shown as small coloured circles. The values measured in GN-2 \citep{Tang2026b} are show as white-filled squares and are placed at the mean corresponding spatial position in the model. The ballistic radial velocity law for different BH masses and Eddington ratios are shown, $M_{\rm BH}(1-\Gamma)$. The velocity law inferred from the model is consistent with $M_{\rm BH}\sim10^6 M_\odot$ and $\Gamma\sim0.9$ (blue line). Independent from, but consistent with, the velocity law, $\Gamma M_{\rm BH}$ is also constrained by the observed luminosity. Lower panel: ion fractions for different sight-lines illustrating spatial stratification.
}
\label{fig:abs_centroidII}
\end{figure}

In high-$z$ N-emitters the N-rich line-emitting gas may be only a small fraction of the ISM \citep[e.g.][]{Marques-Chaves2024}, but here the short depletion time of the broad-line gas means that we can infer the rate at which this small mass is turned over. The mass of CNO-processed material, so-called hydrogen burning `ash', is just $M_{\rm ash} = fM_{\rm gas}$, with a correction factor for the relative metallicity of the cocoon compared to the stars contributing the ash, so $M_{\rm ash} = fM_{\rm gas}\, (Z_{\rm cocoon}/Z_{\rm \star})$ where $Z_{\star}$ is the original metallicity of the polluting stars and by extension of their ash. Since it is very likely that the gas that formed the stars and the nuclear gas share a common origin, $Z_{\rm cocoon}\simeq Z_\star$ unless the nucleus was since diluted by pristine inflow. The many metal features observed in LRDs show it is not pristine and support adopting similar enrichment as the preceding stars. The mass of the instantaneously observed line-emitting cocoon is modest and thus the  N-enhanced ash is also modest, $M_{\rm ash}\!\sim\!10^2\,\mathrm{M_\odot}$ for the inferred $f$ values above. For a yield of ash per unit stellar mass formed of $\eta_{\rm CN}\!=\!0.01$, i.e., the mass of hydrogen-burning processed gas returned by a full stellar population, for which the massive stars dominate \citep{Charbonnel2023,Higgins2025}, this requires only $\sim\!10^{4}\,\mathrm{M_\odot}$ of pre-supernova stars to supply the instantaneous cocoon. But at the observed outflow speeds of a few hundred km/s, this gas is being replaced on $R/v\sim$100 yr timescale in order to be observed \citep{Sneppen2026}. Sustaining observable N-enrichment on timescales, $t_{\rm N}$, would require $M_\star=f\,\dot M\,t_{\rm N}\,(Z_{\rm cocoon}/Z_\star)/\eta_{\rm CN}$, where $\dot M$ is the mass flux leaving the N-rich region. The active LRD phase is not measured directly, but likely of order $\sim\!10\,\rm {Myr}$ given how common the LRD phenomenon is \citep{Sun2026}, while the characteristic outflows from \sirocco carry masses of $\dot M_{\rm wind}\simeq O(1\,\mathrm{M_\odot}$\,yr$^{-1})$ in \citet{Sneppen2026}. If the wind escapes, $M_{\rm\star,esc}\,(\eta_{\rm CN}/0.01)\!\sim\!10^{9}\,M_\odot \left({\dot M_{\rm wind}}/{M_\odot\,{\rm yr^{-1}}}\right)$ of stars are needed within the central parsecs. If, however, the gas is recycled, the requisite mass budget is no longer the integral of the wind's mass-loss rate, but now only needs to match the much smaller accreted stream, $\dot M_{\rm BH}\propto\Gamma L_{\rm Edd}$ and $M_{\star,\rm rec} \left({\eta_{\rm CN}}/{0.01}\right)\!\sim\!10^{7} M_\odot \left({\Gamma M_{BH}}/{10^{6}M_\odot}\right)$. A partial escape fraction, $f_{\rm esc}$, lies somewhere between the two extremes:
\begin{equation}
\begin{aligned}
M_\star\,\left(\frac{\eta_{\rm CN}}{0.01}\right)\simeq\;&
\left[10^{7}\left(\frac{\Gamma M_{\rm BH}}{10^{6}\,M_\odot}\right)
+10^{9}\,f_{\rm esc}\left(\frac{\dot M_{\rm wind}}{M_\odot\,{\rm yr^{-1}}}\right)\right]\\
&\times\left(\frac{f}{0.5}\right)\left(\frac{t_N}{10^{7}\,{\rm yr}}\right)
\left(\frac{Z_{\rm cocoon}}{Z_\star}\right)M_\odot
\end{aligned}
\label{eq:m2}
\end{equation}

For standard stellar populations, the observed UV luminosity \citep[$M_{\rm UV}=-20.7$ for GN-2,][]{Tang2026b,Kokorev2026} sets an independent upper limit on the stellar mass, which is most stringent for the young population implied by the supernova-free abundance pattern. For a \citet{Kroupa2001} IMF and a $\lesssim3$\,Myr stellar burst without dust, \textsc{Bagpipes} \citep{Carnall2018} gives $\lesssim10^{8}\,{\rm M_\odot}$ (see Fig.~\ref{fig:muv}). The limit is insensitive to the star-formation history, because i) continuous star formation over $t_N$ gives the same $M_{\rm UV}$ per unit mass as the 1 Myr old burst and ii) any population old enough to relax it has passed through its supernovae. This limit is conservative in two other respects. First, the prominent absorption \citep[with a different pattern from star-forming galaxies at similar redshifts,][]{Kokorev2026} and the \sirocco modelling here favour a sub-dominant stellar contribution to the UV \citep[however, see discussion in][]{Tang2026b}. Second, stars beyond the central parsecs contribute to $M_{\rm UV}$ without feeding the black hole. Even so, this stellar mass upper bound disfavours a fully escaping wind and instead suggests $f_{\rm esc}\!\lesssim\!0.2$ (see Fig.~\ref{fig:muv}). Polluters with a far higher yield per unit mass, $\eta_{\rm CN}\gtrsim0.5$, evade this bound. For instance, the same escaping wind could be fed by tens of supermassive stars of $\sim\!10^5\,{M_\odot}$ \citep{Gieles2018} or by $\gtrsim10^5$ very massive stars \citep[e.g.\ between 100--300\,M$_\odot$ and $Z\lesssim0.2Z_\odot$ with $\eta_{\rm CN}\lesssim0.5$ per unit VMS mass,][]{Higgins2025}. Gas recycling with the pathway-agnostic Eq.~\ref{eq:m2} can lower these requirements by more than an order-of-magnitude for $f_{\rm esc}\!<\!0.1$. We now turn to an independent argument for gas recycling from absorber kinematics.

\subsection{Metal absorption -- a ballistic velocity law}\label{sec:abs_kin}
\begin{figure}
\vspace{-0.1cm}
\begin{center}
    \includegraphics[angle=0,width=0.45\textwidth]{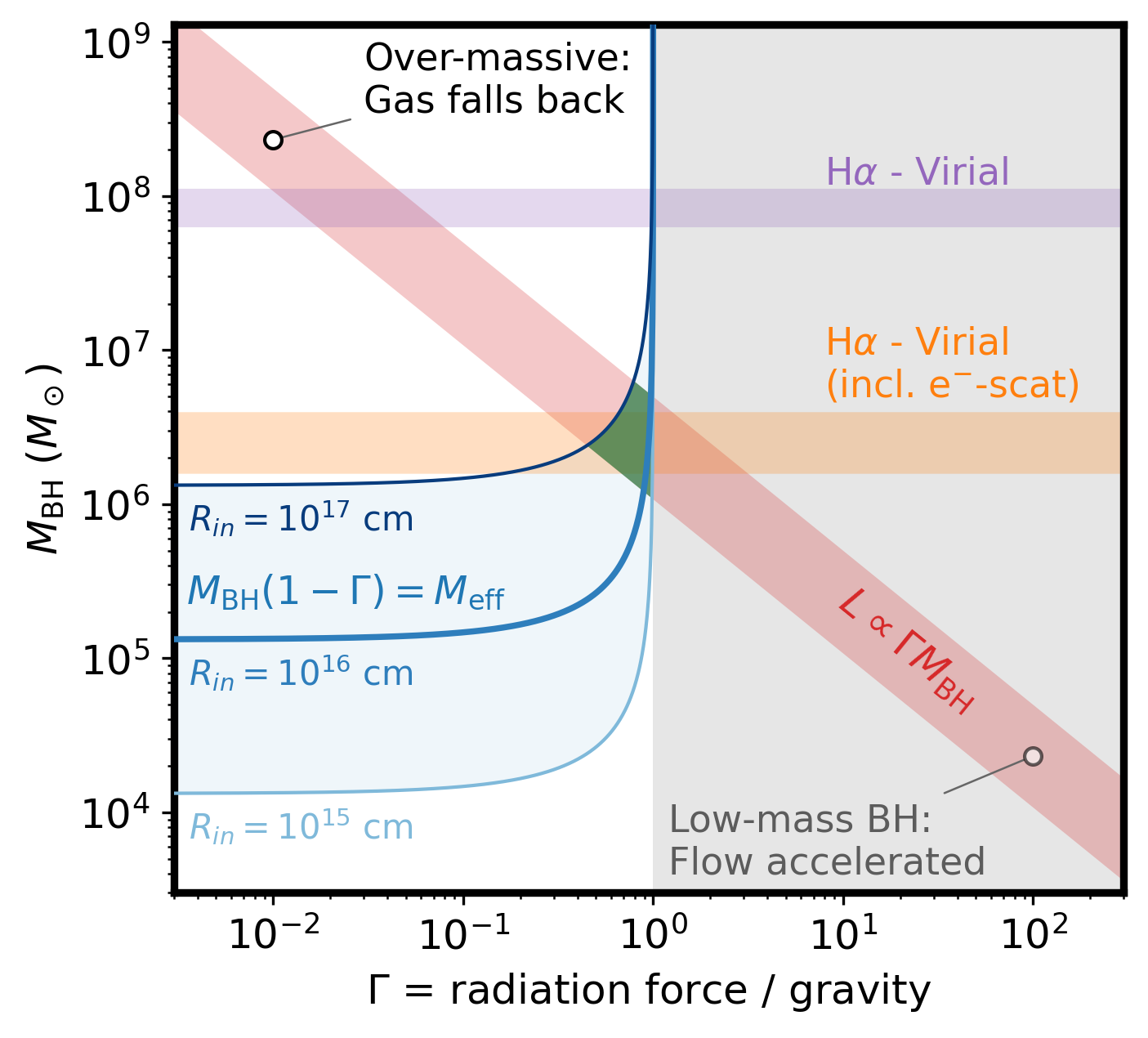}
\end{center}
\vspace{-0.6cm}
\caption{Black hole mass constraints for ballistic wind in GN-2. The illustrative constraints from the bolometric luminosity ($\propto \Gamma M_{\rm BH}$, red), the decelerating gas ($\propto M_{\rm BH} (1-\Gamma)$, blue) and joint constraints (green), imply a limited range of allowed BH masses. Low-mass BHs ($\sim10^4\,M_{\rm \odot}$) become radiation-pressure dominated, while over-massive black holes ($\sim10^8\,M_{\rm \odot}$) would rapidly decelerate the gas. Mass constraints using the single-epoch virial H$\alpha$ estimator based on the full line width (purple) and from the electron-scattering decomposition \citep[orange, following][]{Rusakov2025,Nikopoulos2025,Sneppen2026} calibrated on local AGN \citep[e.g.][]{Reines2015} are shown for comparison. The BH mass constraints from the ballistic wind model are only consistent when electron-scattering broadening is accounted for.}
\label{fig:BH_mass}
\end{figure}

Several, often saturated, narrow absorption features are produced by the metal ions in the enshrouding gas in \sirocco models. Because the ionisation states are radially stratified, a velocity field that changes with radius gives a systematic difference in their line velocity centroids (Fig.~\ref{fig:abs_centroid}). In the models, the velocity gradient of these lines with increasing ionisation state is i) positive for a decelerating flow, indicating a gas outflow coasting in the SMBH gravitational potential or ii) negative for an accelerating flow such as from a radiatively-driven flow. The observations of GN-2 and CEERS-7902 show the former with fast \ion{C}{iv} and slow \ion{Al}{ii} (for the saturated \ion{Mg}{ii}, see App.~\ref{app:abund}). \sirocco reproduces this ordering with a ballistic law in which gas coasts near the local escape speed with an effective gravity $M_{\rm eff} \sim 10^{5}\,\mathrm{M_\odot}$ (see Fig.~\ref{fig:abs_centroidII}). 
The effective mass is not the black-hole mass but the residual of gravity and radiation pressure, $M_{\rm eff}=M_{\rm BH}(1-\Gamma)$, with $\Gamma$ being the ratio of the forces. 
By definition $\Gamma\!\sim\!1$ close to the Eddington accretion limit and indeed is in the \sirocco model in Fig.~\ref{fig:abs_centroidII}, when its own radiation field is used to compute the radiation force. 
For the model in Fig.~\ref{fig:abs_centroidII} with a representative LRD luminosity $L = 10^{44}$\,erg\,s$^{-1}$ this requires $M_{\rm BH} \simeq (1{-}2)\times10^{6}\,\mathrm{M_\odot}$. For comparison, an over-massive SMBH ($\sim10^8\,\mathrm{M_\odot}, \Gamma\sim0.01$), by definition very sub-Eddington at these luminosities, would decelerate the flow immediately, whereas a low-mass black hole (e.g.\ $\sim10^4\,\mathrm{M_\odot}, \Gamma\sim100$) would not decelerate the flow at all and be dominated by the radiation field (see Fig.~\ref{fig:BH_mass}). Within the \sirocco framework, the proposed black hole mass estimate is uncertain to the same level as the bolometric luminosity and is thus likely robust to $\lesssim$1\,dex (see App.~\ref{app:bh}). 
This SMBH mass estimate is significantly below broad line virial estimates for LRDs \citep[e.g.][]{Maiolino2025,DEugenio2025}, but is comparable to virial relation estimates when corrected for electron-scattering broadened lines \citep[e.g.][]{Rusakov2025,Sneppen2026}, and well above the $10^{4-5}\,\mathrm{M_\odot}$ that \citet{Sun2026} infer when treating the blackbody radius of the continuum as a physical pseudo-photosphere. 

The ballistic model suggests the gas is launched close to the local escape speed and thus could remain bound. Such material turns around on a $t=r/v\sim100\,{\rm yr}$ timescale, which is the same timescale on which the line‑emitting cocoon is replaced \citep{Sneppen2026}. It is thus likely the nitrogen‑rich gas is not lost but cycles through the cocoon, which is what Eq.~\ref{eq:m2} assumes, and the comparable inflow and outflow mass fluxes required by the symmetric electron‑scattering wings \citep{Sneppen2026} are the natural signature of such a fountain \citep[see also][]{Kaaz2026}.
We already require, based on line-shape arguments and H and He absorber velocities, that simultaneous inflows of accretion and outflows caused by feedback are present \citep{Sneppen2026}, suggesting that these absorber kinematics are directional. Objects with faster outflows with lower column density will likely preferentially show up in the blue part of the colour-space of gas-enshrouded AGN \citep[e.g.\ Fig.~10 in][]{Mascia2026}. The clearest absorber ordering is seen in the bluest of the LRDs, GN-2 \citep[$\beta_{\rm opt}\sim0.1,\ \beta_{\rm FUV}\sim-2.4$;][]{Tang2026b}, whereas the comparatively redder GN-29 \citep[$\beta_{\rm opt}\sim0.4,\ \beta_{\rm FUV}\sim-1.4$;][]{Tang2026b}, shows every absorber near systemic and has no velocity gradient, as expected for a sight-line not probing the unbound wind. With only four objects this is suggestive, but requires a larger sample spanning the LRD colour sequence.

\begin{figure}
\vspace{-0.2cm}
\begin{center}
    \includegraphics[angle=0,width=0.5\textwidth,viewport=0 0 410 410 ,clip=]{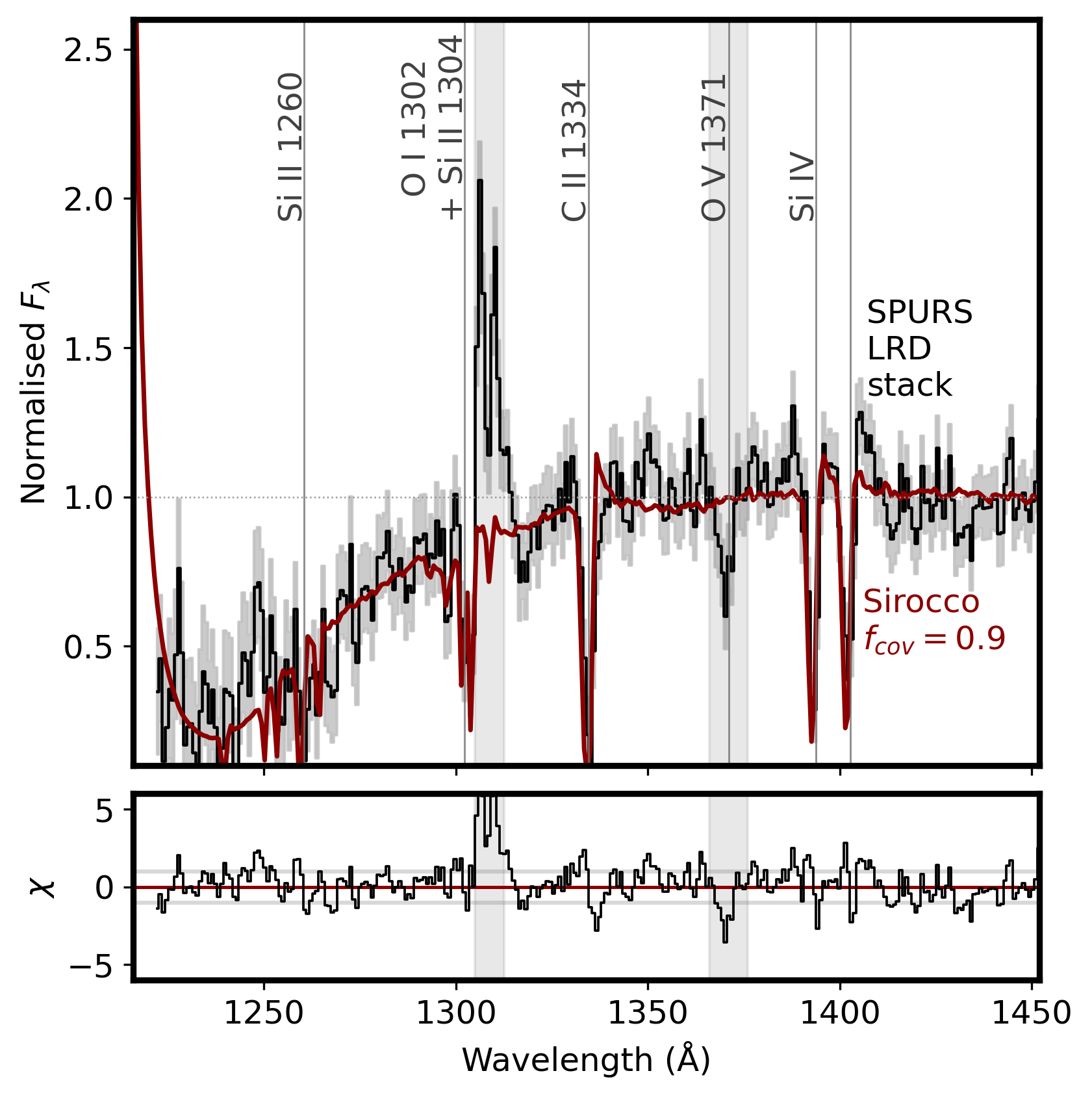}
\end{center}
\vspace{-0.5cm}
\caption{
Stack of the four SPURS LRDs \citep[e.g.\ following Fig.~1 in][]{Kokorev2026} compared to a \sirocco model. Residuals are shown in the lower panel. The \sirocco enshrouding gas is at \(10^{16}\,{\rm cm}{-}10^{19}\,{\rm cm}\), and is largely ionised below $10^{18}\,{\rm cm}$ and neutral above. The model naturally produces strong metal ion absorbers within a more modest $N_{\rm H\,I}$ column due to the much higher column of ionised gas. The $N_{\rm H\,I}$ column from the model is used to add a Ly$\alpha$ absorber wing assuming a large covering fraction ($f_{\rm cov}=0.9$) and not otherwise included in \sirocco. The grey regions mark features that \sirocco does not treat, such as i) \ion{O}{V}\,$\lambda$1371, absorbed from a level 19.7\,eV above the ground state that is populated by trapping in \ion{O}{V}\,$\lambda$630 and ii) the fluorescence lines of \ion{O}{I}$^{*}$ and \ion{Si}{II}$^{*}$, whose re-emission into the fine-structure levels is neglected.  
}
\label{fig:NH}
\end{figure}

\subsection{Abundance inferences from UV lines}
The spatial stratification of different ions also makes it difficult to assess the intrinsic abundance ratio of elements. For example, fitting and interpreting the \ion{Al}{III} absorption line in the model, [Al/Fe] or [Al/Mg] seems 1--2 orders of magnitude supersolar, even though the input metal abundance follows a solar composition (see Fig.~\ref{fig:al}). The recovered ratios become decoupled from the underlying abundances for two reasons. First, higher ionisation is reached deeper within the gas cocoon and thus in a higher density region. Second, the model troughs are often saturated but partially refilled by the diffuse emission from the cocoon itself. Similarly, abundances expressed relative to the neutral gas column from damped Ly$\alpha$ also require this warm and cold stratification to be modelled. In these \sirocco models only the outer cocoon is neutral, so $N_{\rm H\,I}\ll N_{\rm H}$, while Fig.~\ref{fig:NH} illustrates the damping wing from an enshrouding cocoon at \(10^{16}{-}10^{19}\)\,cm, which is largely neutral beyond \(\gtrsim10^{18}\)\,cm and with $\log({N}\rm_{H\,I}/cm^{-2})\simeq22.7$. For such a column the UV is barely reddened, requiring only $A_V=0.5{-}0.9$ for a \citet{Calzetti2000} attenuation law to match the UV-slope from \sirocco model in Fig.~\ref{fig:NH}, so the neutral gas has $N_{\rm H,I}/A_V=5^{+4}_{-2}\times10^{22}$\,cm$^{-2}$\,mag$^{-1}$, 10--30 times the Galactic value \citep{Watson2011}. Such a low dust-to-neutral-gas ratio is an expected consequence of LRDs being in low metallicity systems at $\sim 0.1\,Z_{\odot}$ \citep{Nikopoulos2026,Tang2026b}, and the possibly
linear \citep{Heintz2023,Hamanowicz2024} or super-linear decline seen in local low-metallicity dwarfs \citep{RemuRuyer2014}.

\section{Conclusions}

Deep rest-UV spectroscopy from SPURS resolves the stratified metal absorbers and weak broad emission lines anticipated by the \sirocco cocooned-AGN models. Interpreted within the radiative transfer framework, the data yields the following constraints. 
\begin{enumerate}
    \item The broad \ion{N}{iii}]/\ion{C}{iii}] and \ion{N}{iv}]/\ion{C}{iv} ratios of GN-2 indicate N/C $= 12_{-2}^{+3}\,(\mathrm{N/C})_\odot$ in the broad-line gas, matching its narrow-line values for the same lines and placing the LRD proper among the high-redshift N-emitters. About half or more of CNO nuclei have been through hydrogen burning. Given the short depletion times of the broad-line proper, this provides a progenitor-independent limit for the preceding stellar enrichment. An escaping wind and a sustained nitrogen-rich phase ($t_N\!\sim\!10$ Myr) are not compatible with the $M_{UV}$ stellar‑mass limit. 
    \item The high ratio of \ion{N}{iv}]/\ion{C}{iv} puts an upper bound on the emitting region's electron density, $n_e<2\times10^{10} \,{\rm cm^{-3}}$, which disfavours models with large clumping factors (e.g., by factors $\gtrsim 20$). 
    \item Absorption line velocities are ordered by ionisation potential (e.g.\ fast \ion{C}{iv} and slow \ion{Al}{ii} in GN-2 and CEERS-7902) indicating a velocity law that decreases outward in the models. A ballistic flow coasting near the escape speed reproduces this ordering, and implies $M_\mathrm{BH} \simeq 10^{6-7}\,\mathrm{M_\odot}$ and near-Eddington accretion.
    \item Apparent optical-depth abundances do not reflect the input abundance of the \sirocco models due to spatial stratification and refilled saturated line troughs. A cocoon with solar abundances and saturated lines naturally reproduces Al-rich, Mg-poor absorption pattern reported for LRDs \citep[see][and App.~\ref{app:abund}]{Kokorev2026}. 
   
\end{enumerate}
Recycled gas is favoured by diagnostics in both emission (the nitrogen budget) and absorption (marginally bound, decelerating flow). Thus, the LRD cocoon may be a partially failed wind, where feedback at near‑Eddington luminosity lifts hydrogen‑burnt gas, which falls back and is re‑accreted. Deep rest‑UV spectroscopy of LRDs is still limited to a handful of objects. The far-reaching conclusions that can be drawn in this regime and exemplified here suggest that significantly larger UV observational samples will be important to understand the LRD phenomenon.



\begin{acknowledgements}
We would like to thank Charlotte Mason for insightful comments and suggestions on the manuscript. AS, DW are funded in part by the European Union (ERC, HEAVYMETAL, 101071865). Views and opinions expressed are, however, those of the authors only and do not necessarily reflect those of the European Union or the European Research Council. Neither the European Union nor the granting authority can be held responsible for them. AS and KEH acknowledge support from Independent Research Fund Denmark (DFF) under grant 6219-00033B and 5251-00009B, respectively. The Cosmic Dawn Center (DAWN) is funded by the Danish National Research Foundation under grant DNRF140. JHM acknowledges funding from a Royal Society University Research Fellowship (URFR1221062). 
\end{acknowledgements}

\bibliographystyle{aa}
\bibliography{refs} 

\begin{thebibliography}{88}
\expandafter\ifx\csname natexlab\endcsname\relax\def\natexlab#1{#1}\fi

\bibitem[{{Akins} {et~al.}(2025){Akins}, {Casey}, {Berg}, {Chisholm}, {Cloonan}, {Franco}, {Finkelstein}, {Fujimoto}, {Koekemoer}, {Kokorev}, {Lambrides}, {Robertson}, {Taylor}, {Coulter}, {Fox}, \& {Karmen}}]{Akins2025b}
{Akins}, H.~B., {Casey}, C.~M., {Berg}, D.~A., {et~al.} 2025, \apjl, 980, L29

\bibitem[{{Ando} {et~al.}(2026){Ando}, {Harikane}, {Katz}, {Inayoshi}, \& {Tanaka}}]{Ando2026}
{Ando}, M., {Harikane}, Y., {Katz}, H., {Inayoshi}, K., \& {Tanaka}, T.~S. 2026, arXiv e-prints, arXiv:2606.03522

\bibitem[{Arnould {et~al.}(1999)Arnould, Goriely, \& Jorissen}]{Arnould1999}
Arnould, M., Goriely, S., \& Jorissen, A. 1999, Non-explosive hydrogen and helium burnings: Abundance predictions from the NACRE reaction rate compilation

\bibitem[{{Baldwin} {et~al.}(2003){Baldwin}, {Hamann}, {Korista}, {Ferland}, {Dietrich}, \& {Warner}}]{Baldwin2003}
{Baldwin}, J.~A., {Hamann}, F., {Korista}, K.~T., {et~al.} 2003, \apj, 583, 649

\bibitem[{{Bellovary}(2025)}]{Bellovary2025}
{Bellovary}, J. 2025, arXiv e-prints, arXiv:2501.03309

\bibitem[{{Brazzini} {et~al.}(2026){Brazzini}, {D'Eugenio}, {Maiolino}, {Lyu}, {DeCoursey}, {{\"U}bler}, {Ji}, {Juod{\v{z}}balis}, {Scholtz}, {Jones}, {Hainline}, {Dalla Bont{\`a}}, {{\'e}rez-Gonz{\'a}lez}, {Geris}, {Harshan}, {Feruglio}, {Bischetti}, {Mazzolari}, {Rieke}, {Alberts}, {Trefoloni}, {Carniani}, {Parlanti}, {Marconi}, {Risaliti}, {Ramos Almeida}, {Rinaldi}, {Perna}, {Zamora}, {Lamperti}, {Venturi}, {Cresci}, {Bunker}, \& {Ivey}}]{Brazzini2026}
{Brazzini}, M., {D'Eugenio}, F., {Maiolino}, R., {et~al.} 2026, arXiv e-prints, arXiv:2601.22214

\bibitem[{{Calzetti} {et~al.}(2000){Calzetti}, {Armus}, {Bohlin}, {Kinney}, {Koornneef}, \& {Storchi-Bergmann}}]{Calzetti2000}
{Calzetti}, D., {Armus}, L., {Bohlin}, R.~C., {et~al.} 2000, \apj, 533, 682

\bibitem[{Cameron {et~al.}(2026)Cameron, Carreira, Simmonds, Bunker, Saxena, Carniani, Charlot, Chevallard, Curtis-Lake, Hainline, Hausen, Ji, Ji, Johnson, Rinaldi, Robertson, Scholtz, Silcock, Tacchella, Trussler, Übler, Williams, Willmer, Willott, \& Witstok}]{Cameron2026}
Cameron, A.~J., Carreira, C., Simmonds, C., {et~al.} 2026, Monthly Notices of the Royal Astronomical Society, 552

\bibitem[{{Cameron} {et~al.}(2023){Cameron}, {Katz}, {Rey}, \& {Saxena}}]{Cameron2023}
{Cameron}, A.~J., {Katz}, H., {Rey}, M.~P., \& {Saxena}, A. 2023, \mnras, 523, 3516

\bibitem[{{Cameron} {et~al.}(2024){Cameron}, {Katz}, {Witten}, {Saxena}, {Laporte}, \& {Bunker}}]{Cameron2024}
{Cameron}, A.~J., {Katz}, H., {Witten}, C., {et~al.} 2024, \mnras, 534, 523

\bibitem[{{Carnall} {et~al.}(2018){Carnall}, {McLure}, {Dunlop}, \& {Dav{\'e}}}]{Carnall2018}
{Carnall}, A.~C., {McLure}, R.~J., {Dunlop}, J.~S., \& {Dav{\'e}}, R. 2018, \mnras, 480, 4379

\bibitem[{{Charbonnel} {et~al.}(2023){Charbonnel}, {Schaerer}, {Prantzos}, {Ram{\'\i}rez-Galeano}, {Fragos}, {Kuruvanthodi}, {Marques-Chaves}, \& {Gieles}}]{Charbonnel2023}
{Charbonnel}, C., {Schaerer}, D., {Prantzos}, N., {et~al.} 2023, \aap, 673, L7

\bibitem[{Chen {et~al.}(2026)Chen, Stark, Mason, Plat, Gelli, Senchyna, Endsley, Tang, Topping, Whitler, {et~al.}}]{Chen2026spurs}
Chen, Z., Stark, D.~P., Mason, C.~A., {et~al.} 2026, arXiv preprint arXiv:2604.21516

\bibitem[{{Chen} {et~al.}(2026){Chen}, {Stark}, {Mason}, {Senchyna}, {Tang}, {Keerthi Vasan G.}, {Whitler}, {Plat}, \& {Gelli}}]{Chen2026}
{Chen}, Z., {Stark}, D.~P., {Mason}, C.~A., {et~al.} 2026, arXiv e-prints, arXiv:2608.12699

\bibitem[{{Chisholm} {et~al.}(2026){Chisholm}, {Berg}, {Boylan-Kolchin}, {de Graaff}, {Furtak}, {Kokorev}, {Matthee}, {Mu{\~n}oz}, {Naidu}, \& {Sander}}]{Chisholm2026}
{Chisholm}, J., {Berg}, D.~A., {Boylan-Kolchin}, M., {et~al.} 2026, \apjl, 1004, L4

\bibitem[{{Cloonan} {et~al.}(2026){Cloonan}, {Whitaker}, {Manning}, {Williams}, {Greene}, {Oesch}, {Weibel}, {Brammer}, {de Graaff}, {Hviding}, {Dayal}, {Jespersen}, {Ji}, {Labbe}, {Xiao}, \& {Zhang}}]{Cloonan2026}
{Cloonan}, A.~P., {Whitaker}, K.~E., {Manning}, S.~M., {et~al.} 2026, arXiv e-prints, arXiv:2603.24700

\bibitem[{{Crowther} {et~al.}(2010){Crowther}, {Schnurr}, {Hirschi}, {Yusof}, {Parker}, {Goodwin}, \& {Kassim}}]{Crowther2010}
{Crowther}, P.~A., {Schnurr}, O., {Hirschi}, R., {et~al.} 2010, \mnras, 408, 731

\bibitem[{{de Graaff} {et~al.}(2025{\natexlab{a}}){de Graaff}, {Hviding}, {Naidu}, {Greene}, {Miller}, {Leja}, {Matthee}, {Brammer}, {Katz}, {Bezanson}, {Boogaard}, {Bose}, {Chisholm}, {Cleri}, {Dayal}, {Feldmann}, {Fudamoto}, {Fujimoto}, {Furtak}, {Glazebrook}, {Gottumukkala}, {Heintz}, {Kokorev}, {Labbe}, {Maseda}, {McConachie}, {Nanayakkara}, {Nelson}, {Nowaczyk}, {Oesch}, {Rix}, {Setton}, {Torralba}, {Walter}, {Wang}, {Weibel}, \& {van der Wel}}]{deGraaff2025b}
{de Graaff}, A., {Hviding}, R.~E., {Naidu}, R.~P., {et~al.} 2025{\natexlab{a}}, arXiv e-prints, arXiv:2511.21820

\bibitem[{{de Graaff} {et~al.}(2025{\natexlab{b}}){de Graaff}, {Rix}, {Naidu}, {Labb{\'e}}, {Wang}, {Leja}, {Matthee}, {Katz}, {Greene}, {Hviding}, {Baggen}, {Bezanson}, {Boogaard}, {Brammer}, {Dayal}, {van Dokkum}, {Goulding}, {Hirschmann}, {Maseda}, {McConachie}, {Miller}, {Nelson}, {Oesch}, {Setton}, {Shivaei}, {Weibel}, {Whitaker}, \& {Williams}}]{deGraaff2025}
{de Graaff}, A., {Rix}, H.-W., {Naidu}, R.~P., {et~al.} 2025{\natexlab{b}}, \aap, 701, A168

\bibitem[{{Del Zanna} {et~al.}(2021){Del Zanna}, {Dere}, {Young}, \& {Landi}}]{DelZanna2021}
{Del Zanna}, G., {Dere}, K.~P., {Young}, P.~R., \& {Landi}, E. 2021, The Astrophysical Journal, 909, 38

\bibitem[{{Denissenkov} \& {Hartwick}(2014)}]{Denissenkov2014}
{Denissenkov}, P.~A. \& {Hartwick}, F.~D.~A. 2014, \mnras, 437, L21

\bibitem[{{Dere} {et~al.}(1997){Dere}, {Landi}, {Mason}, {Monsignori Fossi}, \& {Young}}]{Dere1997}
{Dere}, K.~P., {Landi}, E., {Mason}, H.~E., {Monsignori Fossi}, B.~C., \& {Young}, P.~R. 1997, Astronomy and Astrophysics Supplement Series, 125, 149

\bibitem[{{D'Eugenio} {et~al.}(2025){D'Eugenio}, {Maiolino}, {Perna}, {Uebler}, {Ji}, {McClymont}, {Koudmani}, {Sijacki}, {Juod{\v{z}}balis}, {Scholtz}, {Bennett}, {Bunker}, {Carniani}, {Charlot}, {Cresci}, {Curtis-Lake}, {Dalla Bont{\`a}}, {Jones}, {Lyu}, {Marconi}, {Mazzolari}, {Nelson}, {Parlanti}, {Robertson}, {Schneider}, {Simmonds}, {Tacchella}, {Venturi}, {Willott}, {Witstok}, \& {Witten}}]{DEugenio2025}
{D'Eugenio}, F., {Maiolino}, R., {Perna}, M., {et~al.} 2025, arXiv e-prints, arXiv:2503.11752

\bibitem[{{Durodola} {et~al.}(2026){Durodola}, {Hickox}, {Cohn}, \& {Alexander}}]{Durodola2026}
{Durodola}, E., {Hickox}, R.~C., {Cohn}, J., \& {Alexander}, D.~M. 2026, arXiv e-prints, arXiv:2609.13444

\bibitem[{{Geris} {et~al.}(2026){Geris}, {Maiolino}, {Ji}, {Risaliti}, {Lanzuisi}, {D'Eugenio}, {Isobe}, {Jones}, {Harshan}, {Brazzini}, {Juod{\v{z}}balis}, {Scholtz}, {Rinaldi}, {{\"U}bler}, {Baker}, {Bunker}, {Brusa}, {Carniani}, {Charlot}, {Curti}, {Comastri}, {Lake}, {Gilli}, {Hainline}, {Madau}, {Marchesi}, {Mazzolari}, {Napolitano}, {Parlanti}, {Pentericci}, {Ramos Almeida}, {Robertson}, {Silcock}, {Tripodi}, {Venturi}, {Vignali}, {Vito}, \& {Zhu}}]{Geris2026}
{Geris}, S., {Maiolino}, R., {Ji}, X., {et~al.} 2026, arXiv e-prints, arXiv:2606.21614

\bibitem[{{Gieles} {et~al.}(2018){Gieles}, {Charbonnel}, {Krause}, {H{\'e}nault-Brunet}, {Agertz}, {Lamers}, {Bastian}, {Gualandris}, {Zocchi}, \& {Petts}}]{Gieles2018}
{Gieles}, M., {Charbonnel}, C., {Krause}, M. G.~H., {et~al.} 2018, \mnras, 478, 2461

\bibitem[{{Greene} \& {Ho}(2005)}]{Greene2005}
{Greene}, J.~E. \& {Ho}, L.~C. 2005, \apj, 630, 122

\bibitem[{{Greene} {et~al.}(2026){Greene}, {Setton}, {Furtak}, {Naidu}, {Volonteri}, {Dayal}, {Labbe}, {van Dokkum}, {Bezanson}, {Brammer}, {Cutler}, {Glazebrook}, {de Graaff}, {Hirschmann}, {Hviding}, {Kokorev}, {Leja}, {Liu}, {Ma}, {Matthee}, {Nanayakkara}, {Oesch}, {Pan}, {Price}, {Spilker}, {Wang}, {Weaver}, {Whitaker}, {Williams}, \& {Zitrin}}]{Greene2025}
{Greene}, J.~E., {Setton}, D.~J., {Furtak}, L.~J., {et~al.} 2026, \apj, 996, 129

\bibitem[{{Hamanowicz} {et~al.}(2024){Hamanowicz}, {Tchernyshyov}, {Roman-Duval}, {Jenkins}, {Rafelski}, {Gordon}, {Zheng}, {Garcia}, \& {Werk}}]{Hamanowicz2024}
{Hamanowicz}, A., {Tchernyshyov}, K., {Roman-Duval}, J., {et~al.} 2024, \apj, 966, 80

\bibitem[{{Harikane} {et~al.}(2023){Harikane}, {Zhang}, {Nakajima}, {Ouchi}, {Isobe}, {Ono}, {Hatano}, {Xu}, \& {Umeda}}]{Harikane2023}
{Harikane}, Y., {Zhang}, Y., {Nakajima}, K., {et~al.} 2023, \apj, 959, 39

\bibitem[{{Heintz} {et~al.}(2023){Heintz}, {De Cia}, {Th{\"o}ne}, {Krogager}, {Yates}, {Vejlgaard}, {Konstantopoulou}, {Fynbo}, {Watson}, {Narayanan}, {Wilson}, {Arabsalmani}, {Campana}, {D'Elia}, {De Pasquale}, {Hartmann}, {Izzo}, {Jakobsson}, {Kouveliotou}, {Levan}, {Li}, {Malesani}, {Melandri}, {Milvang-Jensen}, {M{\o}ller}, {Palazzi}, {Palmerio}, {Petitjean}, {Pugliese}, {Rossi}, {Saccardi}, {Salvaterra}, {Savaglio}, {Schady}, {Stratta}, {Tanvir}, {de Ugarte Postigo}, {Vergani}, {Wiersema}, {Wijers}, \& {Zafar}}]{Heintz2023}
{Heintz}, K.~E., {De Cia}, A., {Th{\"o}ne}, C.~C., {et~al.} 2023, \aap, 679, A91

\bibitem[{Higgins {et~al.}(2025)Higgins, Vink, Hirschi, Laird, \& Sabhahit}]{Higgins2025}
Higgins, E.~R., Vink, J.~S., Hirschi, R., Laird, A.~M., \& Sabhahit, G. 2025, \aap, 699, A71

\bibitem[{{Inayoshi} \& {Maiolino}(2025)}]{Inayoshi2024}
{Inayoshi}, K. \& {Maiolino}, R. 2025, \apjl, 980, L27

\bibitem[{{Ishikawa} {et~al.}(2026){Ishikawa}, {Eilers}, {Naidu}, {Matthee}, {Bordoloi}, {Chisholm}, {Greene}, {Ma}, {Oesch}, {Sun}, {Torralba}, {Weaver}, {Wuyts}, \& {Xiao}}]{Ishikawa2026}
{Ishikawa}, Y., {Eilers}, A.-C., {Naidu}, R.~P., {et~al.} 2026, arXiv e-prints, arXiv:2607.09647

\bibitem[{{Ji} {et~al.}(2026{\natexlab{a}}){Ji}, {Belokurov}, {Maiolino}, {Monty}, {Isobe}, {Kravtsov}, {McClymont}, \& {{\"U}bler}}]{Ji2026_N}
{Ji}, X., {Belokurov}, V., {Maiolino}, R., {et~al.} 2026{\natexlab{a}}, \mnras, 545, staf2110

\bibitem[{{Ji} {et~al.}(2026{\natexlab{b}}){Ji}, {Pezzulli}, {D'Eugenio}, {Maiolino}, {Carniani}, {Tacchella}, {Jones}, {Smith}, {Witstok}, {Fabian}, {Geris}, {Harshan}, {Isobe}, {Ivey}, {Juod{\v{z}}balis}, {Pascalau}, {Scholtz}, \& {Witten}}]{Ji2026}
{Ji}, X., {Pezzulli}, G., {D'Eugenio}, F., {et~al.} 2026{\natexlab{b}}, arXiv e-prints, arXiv:2604.03370

\bibitem[{{Ji} {et~al.}(2026{\natexlab{c}}){Ji}, {Sun}, {Giavalisco}, {de Graaff}, {Williams}, {Zhu}, {Rieke}, \& {Rieke}}]{Ji2026b}
{Ji}, Z., {Sun}, Y., {Giavalisco}, M., {et~al.} 2026{\natexlab{c}}, arXiv e-prints, arXiv:2606.09970

\bibitem[{Kaaz {et~al.}(2026)Kaaz, Quataert, Govreen-Segal, \& Liu}]{Kaaz2026}
Kaaz, N., Quataert, E., Govreen-Segal, T., \& Liu, H. 2026, Little Red Dots As Super-Eddington Fountain Flows

\bibitem[{{Kageura} {et~al.}(2026){Kageura}, {Ouchi}, {Yanagisawa}, {Ando}, {Harikane}, {Kiyota}, {Nakane}, {Ono}, \& {Takeda}}]{Kageura2026}
{Kageura}, Y., {Ouchi}, M., {Yanagisawa}, H., {et~al.} 2026, arXiv e-prints, arXiv:2608.14534

\bibitem[{{Keerthi Vasan G.} {et~al.}(2026){Keerthi Vasan G.}, {Senchyna}, {Mason}, {Chen}, {Stark}, {Jones}, {Whitler}, {Glazer}, {Aravena}, {Gonzalez-Lopez}, {Endsley}, {Gelli}, {Tang}, \& {Topping}}]{Vasan2026}
{Keerthi Vasan G.}, C., {Senchyna}, P., {Mason}, C.~A., {et~al.} 2026, arXiv e-prints, arXiv:2606.06609

\bibitem[{{Kobayashi} \& {Ferrara}(2024)}]{Kobayashi2024}
{Kobayashi}, C. \& {Ferrara}, A. 2024, \apjl, 962, L6

\bibitem[{{Kocevski} {et~al.}(2023){Kocevski}, {Onoue}, {Inayoshi}, {Trump}, {Arrabal Haro}, {Grazian}, {Dickinson}, {Finkelstein}, {Kartaltepe}, {Hirschmann}, {Aird}, {Holwerda}, {Fujimoto}, {Juneau}, {Amor{\'\i}n}, {Backhaus}, {Bagley}, {Barro}, {Bell}, {Bisigello}, {Calabr{\`o}}, {Cleri}, {Cooper}, {Ding}, {Grogin}, {Ho}, {Hutchison}, {Inoue}, {Jiang}, {Jones}, {Koekemoer}, {Li}, {Li}, {McGrath}, {Molina}, {Papovich}, {P{\'e}rez-Gonz{\'a}lez}, {Pirzkal}, {Wilkins}, {Yang}, \& {Yung}}]{Kocevski2023}
{Kocevski}, D.~D., {Onoue}, M., {Inayoshi}, K., {et~al.} 2023, \apjl, 954, L4

\bibitem[{{Kokorev} {et~al.}(2026){Kokorev}, {Chisholm}, {Naidu}, {Gieles}, {Finkelstein}, {Berg}, {Akins}, {Taylor}, {Fujimoto}, {Furtak}, {Greene}, {de Graaff}, {Hawkins}, {Hsiao}, {Nandal}, {Matthee}, {Monty}, {Rinaldi}, \& {Boylan-Kolchin}}]{Kokorev2026}
{Kokorev}, V., {Chisholm}, J., {Naidu}, R.~P., {et~al.} 2026, arXiv e-prints, arXiv:2609.09271

\bibitem[{{Kroupa}(2001)}]{Kroupa2001}
{Kroupa}, P. 2001, \mnras, 322, 231

\bibitem[{{Labbe} {et~al.}(2024){Labbe}, {Greene}, {Matthee}, {Treiber}, {Kokorev}, {Miller}, {Kramarenko}, {Setton}, {Ma}, {Goulding}, {Bezanson}, {Naidu}, {Williams}, {Atek}, {Brammer}, {Cutler}, {Chemerynska}, {Cloonan}, {Dayal}, {de Graaff}, {Fudamoto}, {Fujimoto}, {Furtak}, {Glazebrook}, {Heintz}, {Leja}, {Marchesini}, {Nanayakkara}, {Nelson}, {Oesch}, {Pan}, {Price}, {Shivaei}, {Sobral}, {Suess}, {van Dokkum}, {Wang}, {Weaver}, {Whitaker}, \& {Zitrin}}]{Labbe2024}
{Labbe}, I., {Greene}, J.~E., {Matthee}, J., {et~al.} 2024, arXiv e-prints, arXiv:2412.04557

\bibitem[{{Lambrides} {et~al.}(2026){Lambrides}, {Larson}, {Garofali}, {Ptak}, {Chiaberge}, {Long}, {Hutchison}, {Norman}, {McKinney}, {Akins}, {Berg}, {Chisholm}, {Civano}, {Cloonan}, {Endsley}, {Faisst}, {Gilli}, {Gillman}, {Hirschmann}, {Kartaltepe}, {Kocevski}, {Kokorev}, {Pacucci}, {Richardson}, {Stiavelli}, \& {Whalen}}]{Lambrides2024}
{Lambrides}, E., {Larson}, R.~L., {Garofali}, K., {et~al.} 2026, Nature Astronomy [\eprint[arXiv]{2409.13047}]

\bibitem[{{Li} {et~al.}(2026{\natexlab{a}}){Li}, {Berg}, {Gazagnes}, {Heckman}, {Gronke}, {Erb}, \& {Chisholm}}]{Li2026a}
{Li}, Z., {Berg}, D.~A., {Gazagnes}, S., {et~al.} 2026{\natexlab{a}}, arXiv e-prints, arXiv:2609.11497

\bibitem[{{Li} {et~al.}(2026{\natexlab{b}}){Li}, {Heckman}, {Gronke}, {Xu}, {Henry}, {Schneider}, {Abruzzo}, {Berg}, {James}, {Martin}, \& {Chisholm}}]{Li2026}
{Li}, Z., {Heckman}, T., {Gronke}, M., {et~al.} 2026{\natexlab{b}}, arXiv e-prints, arXiv:2603.06546

\bibitem[{Lim {et~al.}(2026)Lim, Chang, Gronke, Battaia, Lee, \& Seon}]{Lim2026}
Lim, J., Chang, S.-J., Gronke, M., {et~al.} 2026, Tracing Warm Gas through C IV Radiative Transfer

\bibitem[{{Long} \& {Knigge}(2002)}]{long2002}
{Long}, K.~S. \& {Knigge}, C. 2002, \apj, 579, 725

\bibitem[{{Maiolino} {et~al.}(2026){Maiolino}, {{\"U}bler}, {D'Eugenio}, {Scholtz}, {Juod{\v{z}}balis}, {Ji}, {Perna}, {Bromm}, {Dayal}, {Koudmani}, {Liu}, {Schneider}, {Sijacki}, {Valiante}, {Trinca}, {Zhang}, {Volonteri}, {Inayoshi}, {Carniani}, {Nakajima}, {Isobe}, {Witstok}, {Jones}, {Tacchella}, {Arribas}, {Bunker}, {Cataldi}, {Charlot}, {Curti}, {Fabian}, {Katz}, {Kumari}, {Laporte}, {Mazzolari}, {Robertson}, {Sun}, {Rodriguez Del Pino}, \& {Venturi}}]{Maiolino2025}
{Maiolino}, R., {{\"U}bler}, H., {D'Eugenio}, F., {et~al.} 2026, \mnras, 548, staf2109

\bibitem[{{Marques-Chaves} {et~al.}(2024){Marques-Chaves}, {Schaerer}, {Kuruvanthodi}, {Korber}, {Prantzos}, {Charbonnel}, {Weibel}, {Izotov}, {Messa}, {Brammer}, {Dessauges-Zavadsky}, \& {Oesch}}]{Marques-Chaves2024}
{Marques-Chaves}, R., {Schaerer}, D., {Kuruvanthodi}, A., {et~al.} 2024, \aap, 681, A30

\bibitem[{{Mascia} {et~al.}(2026){Mascia}, {Matthee}, {Torralba}, {Greene}, {Eilers}, {Iani}, \& {Naidu}}]{Mascia2026}
{Mascia}, S., {Matthee}, J., {Torralba}, A., {et~al.} 2026, arXiv e-prints, arXiv:2608.25021

\bibitem[{{Matsuoka} {et~al.}(2017){Matsuoka}, {Nagao}, {Maiolino}, {Marconi}, {Park}, \& {Taniguchi}}]{Matsuoka2017}
{Matsuoka}, K., {Nagao}, T., {Maiolino}, R., {et~al.} 2017, \aap, 608, A90

\bibitem[{{Matthee} {et~al.}(2024){Matthee}, {Naidu}, {Brammer}, {Chisholm}, {Eilers}, {Goulding}, {Greene}, {Kashino}, {Labbe}, {Lilly}, {Mackenzie}, {Oesch}, {Weibel}, {Wuyts}, {Xiao}, {Bordoloi}, {Bouwens}, {van Dokkum}, {Illingworth}, {Kramarenko}, {Maseda}, {Mason}, {Meyer}, {Nelson}, {Reddy}, {Shivaei}, {Simcoe}, \& {Yue}}]{Matthee2024}
{Matthee}, J., {Naidu}, R.~P., {Brammer}, G., {et~al.} 2024, \apj, 963, 129

\bibitem[{{Matthee} {et~al.}(2026){Matthee}, {Torralba}, {Pezzulli}, {Naidu}, {Chisholm}, {Mascia}, {Greene}, {Ishikawa}, {Gronke}, {Wuyts}, {Bordoloi}, {Brammer}, {Chang}, {Eilers}, {de Graaff}, {Hviding}, {Iani}, {Illingworth}, {Kashino}, {Labbe}, {Ma}, {Maseda}, {Meyer}, {Nelson}, {Oesch}, \& {Xiao}}]{Matthee2026}
{Matthee}, J., {Torralba}, A., {Pezzulli}, G., {et~al.} 2026, arXiv e-prints, arXiv:2603.17667

\bibitem[{{Matthews} {et~al.}(2025){Matthews}, {Long}, {Knigge}, {Sim}, {Parkinson}, {Higginbottom}, {Mangham}, {Scepi}, {Wallis}, {Hewitt}, \& {Mosallanezhad}}]{Matthews2025}
{Matthews}, J.~H., {Long}, K.~S., {Knigge}, C., {et~al.} 2025, \mnras, 536, 879

\bibitem[{{Morishita} {et~al.}(2026){Morishita}, {Stiavelli}, {Mason}, {Tripodi}, {Chiaberge}, {Schuldt}, {Willott}, \& {Zhang}}]{Morishita2026}
{Morishita}, T., {Stiavelli}, M., {Mason}, C.~A., {et~al.} 2026, \apj, 998, 175

\bibitem[{{Morton}(2003)}]{Morton2003}
{Morton}, D.~C. 2003, \apjs, 149, 205

\bibitem[{{Naidu} {et~al.}(2025){Naidu}, {Matthee}, {Katz}, {de Graaff}, {Oesch}, {Smith}, {Greene}, {Brammer}, {Weibel}, {Hviding}, {Chisholm}, {Labb{\'e}}, {Simcoe}, {Witten}, {Atek}, {Baggen}, {Belli}, {Bezanson}, {Boogaard}, {Bose}, {Covelo-Paz}, {Dayal}, {Fudamoto}, {Furtak}, {Giovinazzo}, {Goulding}, {Gronke}, {Heintz}, {Hirschmann}, {Illingworth}, {Inoue}, {Johnson}, {Leja}, {Leonova}, {McConachie}, {Maseda}, {Natarajan}, {Nelson}, {Setton}, {Shivaei}, {Sobral}, {Stefanon}, {Tacchella}, {Toft}, {Torralba}, {van Dokkum}, {van der Wel}, {Volonteri}, {Walter}, {Wang}, \& {Watson}}]{Naidu2025}
{Naidu}, R.~P., {Matthee}, J., {Katz}, H., {et~al.} 2025, arXiv e-prints, arXiv:2503.16596

\bibitem[{{Naidu} {et~al.}(2026){Naidu}, {Oesch}, {Brammer}, {Weibel}, {Li}, {Matthee}, {Chisolm}, {Pollock}, {Heintz}, {Johnson}, {Shen}, {Hviding}, {Leja}, {Tacchella}, {Ganguly}, {Witten}, {Atek}, {Belli}, {Bose}, {Bouwens}, {Dayal}, {Decarli}, {de Graaff}, {Fudamoto}, {Giovinazzo}, {Greene}, {Illingworth}, {Inoue}, {Kane}, {Labbe}, {Leonova}, {Marques-Chaves}, {Meyer}, {Nelson}, {Roberts-Borsani}, {Schaerer}, {Simcoe}, {Stefanon}, {Sugahara}, {Toft}, {van der Wel}, {van Dokkum}, {Walter}, {Watson}, {Weaver}, \& {Whitaker}}]{Naidu2026}
{Naidu}, R.~P., {Oesch}, P.~A., {Brammer}, G., {et~al.} 2026, The Open Journal of Astrophysics, 9, 56033

\bibitem[{{Nicholls} {et~al.}(2017){Nicholls}, {Sutherland}, {Dopita}, {Kewley}, \& {Groves}}]{Nicholls2017}
{Nicholls}, D.~C., {Sutherland}, R.~S., {Dopita}, M.~A., {Kewley}, L.~J., \& {Groves}, B.~A. 2017, \mnras, 466, 4403

\bibitem[{{Nikopoulos} {et~al.}(2026{\natexlab{a}}){Nikopoulos}, {Watson}, {Pollock}, {Sneppen}, {Heintz}, {Witstok}, \& {Brammer}}]{Nikopoulos2026}
{Nikopoulos}, G.~P., {Watson}, D., {Pollock}, C.~L., {et~al.} 2026{\natexlab{a}}, arXiv e-prints, arXiv:2606.31515

\bibitem[{{Nikopoulos} {et~al.}(2026{\natexlab{b}}){Nikopoulos}, {Watson}, {Sneppen}, {Rusakov}, {Heintz}, {Witstok}, \& {Brammer}}]{Nikopoulos2025}
{Nikopoulos}, G.~P., {Watson}, D., {Sneppen}, A., {et~al.} 2026{\natexlab{b}}, \aap, 710, A136

\bibitem[{{Papovich} {et~al.}(2026){Papovich}, {Ronayne}, {Hu}, {Kocevski}, {Arrabal Haro}, {Hutchison}, {Finkelstein}, {Lambrides}, {Larson}, {Backhaus}, {Bagley}, {Cleri}, {Dickinson}, {Garner}, {Kartaltepe}, {Kokorev}, {Olivier}, {Taylor}, {Trump}, \& {Yang}}]{Papovich2026}
{Papovich}, C., {Ronayne}, K., {Hu}, W., {et~al.} 2026, arXiv e-prints, arXiv:2608.04101

\bibitem[{{P{\'e}rez-Gonz{\'a}lez} {et~al.}(2026){P{\'e}rez-Gonz{\'a}lez}, {Barro}, {Carniani}, {D'Eugenio}, {Rieke}, {Tripodi}, {Bunker}, {Ji}, {Marques-Chaves}, {Schaerer}, {Venturi}, {Ar{\'e}valo-Gonz{\'a}lez}, {Arribas}, {Rinaldi}, {Rodr{\'\i}guez Del Pino}, {Witstok}, {Bhatawdekar}, {Boogaard}, {Charlot}, {Chevallard}, {Costantin}, {Curti}, {Curtis-Lake}, {Daddi}, {Davis}, {Dickinson}, {Donnan}, {Donnan}, {Dunlop}, {Eisenstein}, {Ferguson}, {Fern{\'a}ndez Aranda}, {Finkelstein}, {Fujimoto}, {Gandolfi}, {Giavalisco}, {Grogin}, {Hamed}, {Hirschmann}, {Kartaltepe}, {Kocevski}, {Koekemoer}, {Leung}, {Lofaro}, {Lucas}, {McLeod}, {Melinder}, {{\"O}stlin}, {Papovich}, {Pentericci}, {P{\'e}rez-D{\'\i}az}, {Rieke}, {Scholtz}, {Somerville}, {Stanton}, {Stevenson}, {Shivaei}, {Tacchella}, {Trump}, {{\"U}bler}, {Wang}, {Williams}, {Willmer}, {Yung}, \& {Zhu}}]{PerezGonzalez2026}
{P{\'e}rez-Gonz{\'a}lez}, P.~G., {Barro}, G., {Carniani}, S., {et~al.} 2026, arXiv e-prints, arXiv:2602.20247

\bibitem[{{Reines} \& {Volonteri}(2015)}]{Reines2015}
{Reines}, A.~E. \& {Volonteri}, M. 2015, \apj, 813, 82

\bibitem[{{R{\'e}my-Ruyer} {et~al.}(2014){R{\'e}my-Ruyer}, {Madden}, {Galliano}, {Galametz}, {Takeuchi}, {Asano}, {Zhukovska}, {Lebouteiller}, {Cormier}, {Jones}, {Bocchio}, {Baes}, {Bendo}, {Boquien}, {Boselli}, {DeLooze}, {Doublier-Pritchard}, {Hughes}, {Karczewski}, \& {Spinoglio}}]{RemuRuyer2014}
{R{\'e}my-Ruyer}, A., {Madden}, S.~C., {Galliano}, F., {et~al.} 2014, \aap, 563, A31

\bibitem[{{Rinaldi} {et~al.}(2025){Rinaldi}, {Bonaventura}, {Rieke}, {Alberts}, {Caputi}, {Baker}, {Baum}, {Bhatawdekar}, {Bunker}, {Carniani}, {Curtis-Lake}, {D'Eugenio}, {Egami}, {Ji}, {Johnson}, {Hainline}, {Helton}, {Lin}, {Lyu}, {Ma}, {Maiolino}, {P{\'e}rez-Gonz{\'a}lez}, {Rieke}, {Robertson}, {Shivaei}, {Stone}, {Sun}, {Tacchella}, {{\"U}bler}, {Williams}, {Willmer}, {Willott}, {Zhang}, \& {Zhu}}]{Rinaldi2024}
{Rinaldi}, P., {Bonaventura}, N., {Rieke}, G.~H., {et~al.} 2025, \apj, 992, 71

\bibitem[{{Rusakov} {et~al.}(2026{\natexlab{a}}){Rusakov}, {Conselice}, {Harvey}, {D'Silva}, \& {Austin}}]{Rusakov2026b}
{Rusakov}, V., {Conselice}, C.~J., {Harvey}, T., {D'Silva}, J. C.~J., \& {Austin}, D. 2026{\natexlab{a}}, \mnras [\eprint[arXiv]{2606.18334}]

\bibitem[{{Rusakov} {et~al.}(2026{\natexlab{b}}){Rusakov}, {Watson}, {Nikopoulos}, {Brammer}, {Gottumukkala}, {Harvey}, {Heintz}, {Damgaard}, {Sim}, {Sneppen}, {Vijayan}, {Adams}, {Austin}, {Conselice}, {Goolsby}, {Toft}, \& {Witstok}}]{Rusakov2025}
{Rusakov}, V., {Watson}, D., {Nikopoulos}, G.~P., {et~al.} 2026{\natexlab{b}}, \nat, 649, 574

\bibitem[{{Schaerer} {et~al.}(2024){Schaerer}, {Marques-Chaves}, {Xiao}, \& {Korber}}]{Schaerer2024}
{Schaerer}, D., {Marques-Chaves}, R., {Xiao}, M., \& {Korber}, D. 2024, \aap, 687, L11

\bibitem[{{Setton} {et~al.}(2025){Setton}, {Greene}, {de Graaff}, {Ma}, {Leja}, {Matthee}, {Bezanson}, {Boogaard}, {Cleri}, {Katz}, {Labbe}, {Maseda}, {McConachie}, {Miller}, {Price}, {Suess}, {van Dokkum}, {Wang}, {Weibel}, {Whitaker}, \& {Williams}}]{Setton2024}
{Setton}, D.~J., {Greene}, J.~E., {de Graaff}, A., {et~al.} 2025, \apj, 995, 118

\bibitem[{{Sneppen} {et~al.}(2026{\natexlab{a}}){Sneppen}, {Matthews}, {Watson}, {Cameron}, {Sim}, {Witstok}, {Brammer}, {Heintz}, \& {Nikopoulos}}]{Sneppen2026b}
{Sneppen}, A., {Matthews}, J.~H., {Watson}, D., {et~al.} 2026{\natexlab{a}}, arXiv e-prints, arXiv:2604.09399

\bibitem[{{Sneppen} {et~al.}(2026{\natexlab{b}}){Sneppen}, {Watson}, {Matthews}, \& {Nikopoulos}}]{Sneppen2026d}
{Sneppen}, A., {Watson}, D., {Matthews}, J.~H., \& {Nikopoulos}, G. 2026{\natexlab{b}}, \aap, 714, A465

\bibitem[{{Sneppen} {et~al.}(2026{\natexlab{c}}){Sneppen}, {Watson}, {Matthews}, {Nikopoulos}, {Allen}, {Brammer}, {Damgaard}, {Heintz}, {Knigge}, {Long}, {Rusakov}, {Sim}, \& {Witstok}}]{Sneppen2026}
{Sneppen}, A., {Watson}, D., {Matthews}, J.~H., {et~al.} 2026{\natexlab{c}}, arXiv e-prints, arXiv:2601.18864

\bibitem[{{Sneppen} {et~al.}(2026{\natexlab{d}}){Sneppen}, {Watson}, {Matthews}, \& {Sim}}]{Sneppen2026c}
{Sneppen}, A., {Watson}, D., {Matthews}, J.~H., \& {Sim}, S.~A. 2026{\natexlab{d}}, \aap, 712, A61

\bibitem[{{Sun} {et~al.}(2026){Sun}, {Naidu}, {Matthee}, {de Graaff}, {Chisholm}, {Greene}, {Oesch}, {Torralba}, {Hviding}, {Brammer}, {Simcoe}, {Bose}, {Bouwens}, {Dayal}, {Eilers}, {Fei}, {Furtak}, {Gottumukkala}, {Goulding}, {Heintz}, {Hirschmann}, {Kokorev}, {Leja}, {Liu}, {Natarajan}, {Santarelli}, {Setton}, {Smith}, {Tacchella}, {Volonteri}, {Walter}, {Weibel}, \& {Williams}}]{Sun2026}
{Sun}, W.~Q., {Naidu}, R.~P., {Matthee}, J., {et~al.} 2026, {Little Red Dot {\ensuremath{-}} Host Galaxy = Black Hole Star: A Gas-Enshrouded Heart at the Center of Every Little Red Dot}

\bibitem[{{Tang} {et~al.}(2026{\natexlab{a}}){Tang}, {Stark}, {Mason}, {Chen}, {Jones}, {Searle Grannis}, {Senchyna}, {Whitler}, {Keerthi Vasan G.}, \& {Gelli}}]{Tang2026b}
{Tang}, M., {Stark}, D.~P., {Mason}, C.~A., {et~al.} 2026{\natexlab{a}}, arXiv e-prints, arXiv:2609.11094

\bibitem[{{Tang} {et~al.}(2026{\natexlab{b}}){Tang}, {Stark}, {Mason}, {Chen}, {Katz}, {Gronke}, {Furtak}, {Chang}, {Matthee}, {Whitler}, {Zitrin}, {Endsley}, {Gelli}, {Roychowdhury}, {Senchyna}, {Topping}, \& {Zhang}}]{Tang2026}
{Tang}, M., {Stark}, D.~P., {Mason}, C.~A., {et~al.} 2026{\natexlab{b}}, arXiv e-prints, arXiv:2604.03563

\bibitem[{{Tang} {et~al.}(2025){Tang}, {Stark}, {Plat}, {Feltre}, {Katz}, {Senchyna}, {Mason}, {Whitler}, {Chen}, \& {Topping}}]{Tang2025}
{Tang}, M., {Stark}, D.~P., {Plat}, A., {et~al.} 2025, \apj, 991, 217

\bibitem[{{Topping} {et~al.}(2025){Topping}, {Stark}, {Senchyna}, {Chen}, {Zitrin}, {Endsley}, {Charlot}, {Furtak}, {Maseda}, {Plat}, {Smit}, {Mainali}, {Chevallard}, {Molyneux}, \& {Rigby}}]{Topping2025}
{Topping}, M.~W., {Stark}, D.~P., {Senchyna}, P., {et~al.} 2025, \apj, 980 [\eprint[arXiv]{2407.19009}]

\bibitem[{{Topping} {et~al.}(2024){Topping}, {Stark}, {Senchyna}, {Plat}, {Zitrin}, {Endsley}, {Charlot}, {Furtak}, {Maseda}, {Smit}, {Mainali}, {Chevallard}, {Molyneux}, \& {Rigby}}]{Topping2024}
{Topping}, M.~W., {Stark}, D.~P., {Senchyna}, P., {et~al.} 2024, \mnras, 529, 3301

\bibitem[{Torralba {et~al.}(2026)Torralba, Matthee, Marques-Chaves, Martins, Navarrete, Pezzulli, Iani, Naidu, Brammer, Cesare, Hutter, Kotiwale, Mackenzie, Mascia, Meyer, \& Schaerer}]{Torralba2025_cola}
Torralba, A., Matthee, J., Marques-Chaves, R., {et~al.} 2026, A Little Red Dot emerging from a nitrogen-rich UV-luminous galaxy with Very Massive Stars

\bibitem[{{Vink}(2023)}]{Vink2023}
{Vink}, J.~S. 2023, \aap, 679, L9

\bibitem[{{Wang} {et~al.}(2026){Wang}, {Leja}, {Katz}, {Inayoshi}, {Cleri}, {de Graaff}, {Hviding}, {van Dokkum}, {Greene}, {Labb{\'e}}, {Matthee}, {McConachie}, {Naidu}, \& {Nelson}}]{wang2025missing}
{Wang}, B., {Leja}, J., {Katz}, H., {et~al.} 2026, \apj, 1003, 10

\bibitem[{{Watson}(2011)}]{Watson2011}
{Watson}, D. 2011, \aap, 533, A16

\bibitem[{{Yang} {et~al.}(2017){Yang}, {Wang}, {Ferland}, {Dou}, {Zhou}, {Jiang}, \& {Sheng}}]{Yang2017}
{Yang}, C., {Wang}, T., {Ferland}, G.~J., {et~al.} 2017, ApJ, 846, 150

\end{thebibliography}

\begin{appendix}
\setcounter{figure}{0}
\renewcommand{\thefigure}{A.\arabic{figure}}

\onecolumn

\begin{figure*}
\begin{center}
    \includegraphics[angle=0,width=0.99\textwidth]{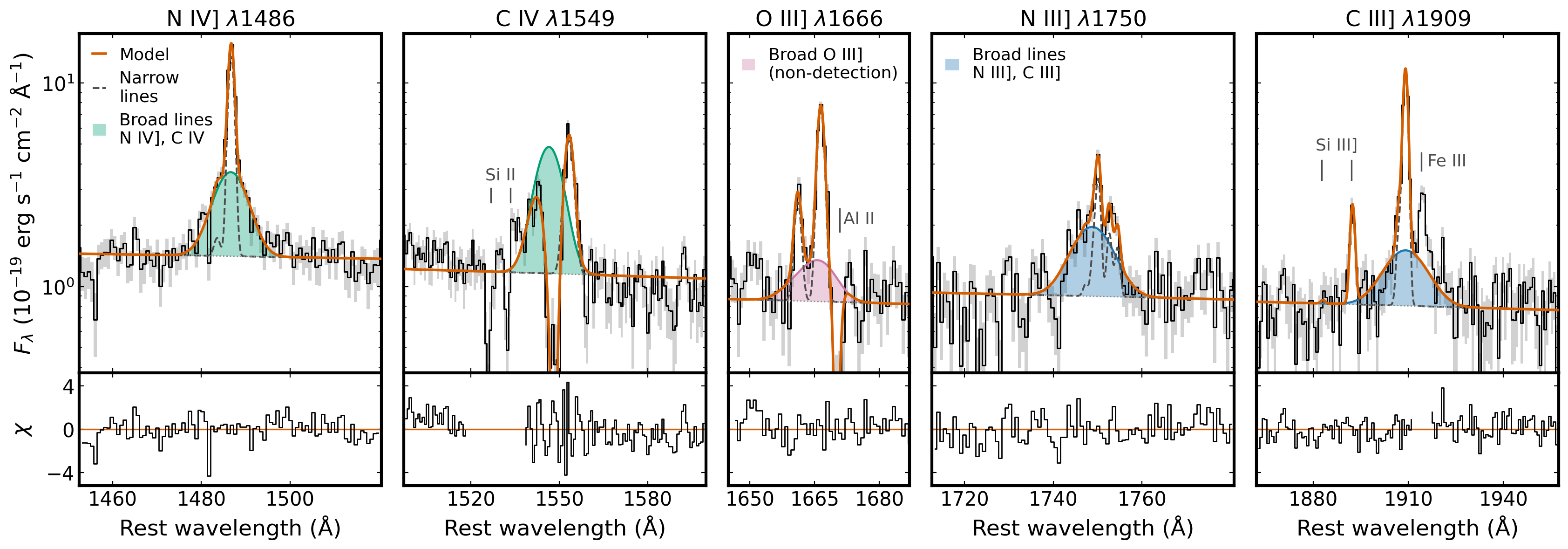}
\end{center}
\vspace{-0.4cm}
\caption{\ion{N}{iv]}, \ion{C}{iv}, \ion{O}{iii]}, \ion{N}{iii]} and \ion{C}{iii]} emission lines observed in GN-2 (black) with best-fit model (orange). The model contains i) a broad Gaussian (shaded), ii) narrow components (dashed line), and iii) a linear continuum (dotted) whose slope is fixed by the line-free continuum. The lower panels highlight the residuals. Neighbouring features are marked (\ion{Si}{II}, \ion{Al}{II}, \ion{Si}{III]}, \ion{Fe}{III}) and excluded from the fits if likely to dominate the line-emission.}
\label{fig:gn2fits}
\end{figure*}


\section{Abundances}\label{app:abund}
Fig.~\ref{fig:nc_density} shows the N/C implied by the broad emission components \ion{N}{iii}]/\ion{C}{iii}] and \ion{N}{iv}]/\ion{C}{iv} ratios of GN-2 \citep{Tang2026b}. Similar $n_e$ and ${\rm N/C}$ constraints follow from our independent fits (see Fig.~\ref{fig:gn2fits}), which give broad-line ratios \ion{N}{IV]}/\ion{C}{IV}$\,=0.52\pm0.14$ and \ion{N}{III]}/\ion{C}{III]}$\,=1.08^{+0.25}_{-0.20}$, against $0.48\pm0.09$ and $1.36\pm0.47$ from \citet{Tang2026b} and 0.058 and 0.064 for solar expectation at $T_{\rm e}=1.5\times10^{4}$\,K and low density ($n_e\lesssim10^{9} \,{\rm cm^{-3}}$). The key assumption behind these constraints is that the ionic fractions of each pair are similar, $x({\rm \ion{N}{III}})\approx x({\rm  \ion{C}{III}})$ and $x({\rm \ion{N}{IV}})\approx x({\rm \ion{C}{IV}})$, as motivated from their nearly coincident ionisation potentials. That the broad components arise in similar gas is supported by the \sirocco models, where $x({\rm \ion{N}{IV}})/x({\rm \ion{C}{IV}})=0.8{-}1$, while $x({\rm \ion{N}{III}})/x({\rm \ion{C}{III}})=0.5{-}0.8$ because \ion{C}{iii} extends to lower ionisation than \ion{N}{iii}. If we adopt these \sirocco ionic ratios, the joint constraint rises on N/C to $\approx13{-}19\,$(N/C)$_\odot$ at $n_e\approx10^{10}$\,cm$^{-3}$, so assuming equal ionic fractions is conservative. 
The computation is evaluated at $T_e=1.5\times10^{4}$\,K, but similar constraints are found for other assumed $T_e=10\,000{-}30\,000$\,K, whereas the N/C confidence intervals are dominated by the observed broad-line ratio uncertainty. At $T_e\gtrsim30\,000$\,K, $n_e$ is no longer constrained from below, but N/C abundance and the upper bound on $n_e$ is a robust feature. Regardless, this is beyond the regime suggested by the \sirocco models for the broad-line emitting gas for these ions. 
Similarly, we explore the sensitivity diagnostics of N\,{\sc iv}]/O\,{\sc iii}]. In the \sirocco models $\sim$80\,\% of O\,{\sc iii}] arises in the N\,{\sc iv}] zone. Notably, if photons capable of ionising to O\,{\sc iv} ($\gtrsim54.9\,$ eV) dominated \ion{N}{iv} formation, prominent broad recombination lines from \ion{He}{iii} ($\gtrsim54.4\,$ eV) would follow, but these recombination lines are weak. For the mixtures that reproduce N/C, CN ash predicts $7{-}15\times10^{-19} \,{\rm erg\,cm^{-2}\,s^{-1}}$ (N/O\,$=4$\,(N/O)$_\odot$) and CNO ash $1{-}5\times10^{-19}$ (13\,(N/O)$_\odot$). The Fig.~\ref{fig:UV} model recomputed with N/O\,$=5$\,(N/O)$_\odot$ gives $6{-}10\times10^{-19}$ once scaled to the observed N\,{\sc iv}]. The bound N/O\,$\gtrsim5$\,(N/O)$_\odot$ spans these uncertainties in ionisation correction, $n_{\rm e}=1{-}20\times10^{9}$\,cm$^{-3}$ and $T_{\rm e}=1{-}3\times10^{4}$\,K. 

To test how reliably elemental abundances can be read from the absorption lines, we applied the apparent-optical-depth recipe of \citet{Kokorev2026} to the emergent \sirocco spectra, smoothed to NIRSpec resolution. Columns of \ion{Mg}{ii}, \ion{Fe}{ii}, \ion{Si}{ii}, \ion{Si}{iv}, \ion{Al}{ii} and \ion{Al}{iii} are converted to [Mg/Fe], [Si/Fe] and [Al/Fe] with their ionisation corrections. All models have solar abundance ratios for these elements, while having different column densities and by extension differing degrees of absorption line saturation. This apparent-optical-depth recipe applied to the \sirocco models returns over-abundant aluminium and often, but not ubiquitously, under-abundant magnesium, close to the values measured in LRDs (Fig.~\ref{fig:al}). This is largely due to two effects. First, \ion{Al}{iii} occupies a deeper and larger column than the low ions, which are confined to the outer ionisation front, so corrections should generally consider and account for stratified gas. Second, the troughs are saturated against the central source but partially refilled by emission from the cocoon itself (Fig.~\ref{fig:al_spec}). 

Because the absorbers are typically saturated, the fitted centroids trace the velocity extent of each ion rather than its column-weighted velocity. Additionally, the trough infilling by emission from greater radii could shift the measured centroids further. However, the comparison uses the same fitting recipe on observations and the emergent model spectra, so these biases likely cancel to first order. For instance, \sirocco \ion{Mg}{ii} shares the gas and mean velocity of the other low ions, yet when the emergent spectrum is fitted the \ion{Mg}{ii} absorber comes out $70$\,km\,s$^{-1}$ bluer than the others, because the large oscillator strength of \ion{Mg}{ii}\,$\lambda2796$ keeps it saturated into the fast tail of the flow. The SPURS LRDs show a \ion{Mg}{ii} offset of similar size and sign, $100\pm40$\,km\,s$^{-1}$ \citep{Tang2026b}, which the models thus reproduce without any kinematic difference between the ions. Thus, we stress the ionisation ordering should be read as an ordering of velocity extents, which depend on line strength as well as on kinematics. 

\setcounter{figure}{0}
\renewcommand{\thefigure}{B.\arabic{figure}}

\begin{figure*}
\begin{center}
    \includegraphics[angle=0,width=\textwidth]{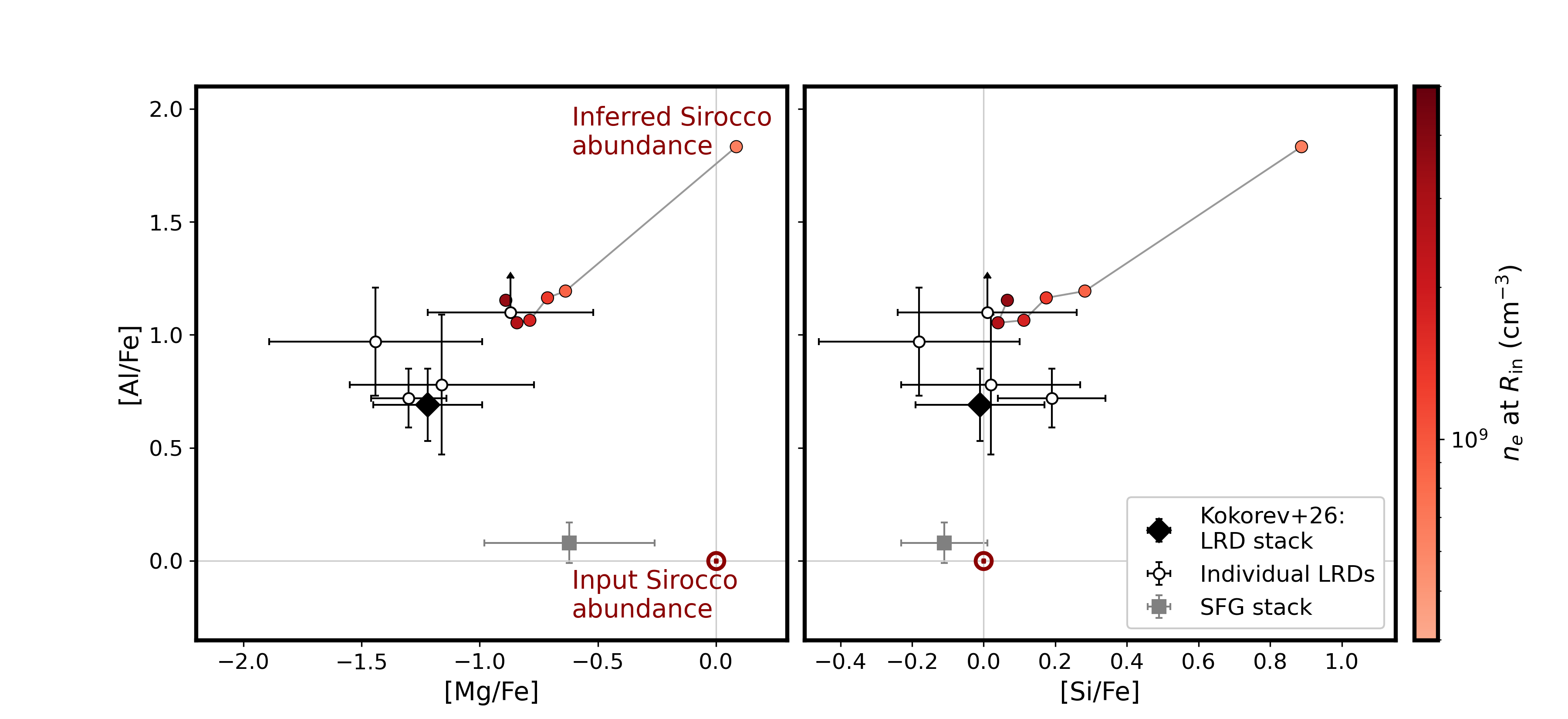}
\end{center}
\vspace{-0.6cm}
\caption{Abundance ratios from the emergent \sirocco spectrum recovered by the apparent-optical-depth method from \citet{Kokorev2026}, versus the SPURS LRD stack and individual LRDs (in black), their star-forming galaxy stack (grey square) and the true, solar-scaled input of every model (red $\odot$). Spatial stratification and saturated, partially refilled troughs bias the models away from the input composition. When only some lines are saturated the bias varies from model to model, but in the high-column, saturated limit [Al/Fe], [Mg/Fe] and [Si/Fe] are close to the region spanned by the individual LRDs. }
\label{fig:al}
\end{figure*}

\begin{figure*}
\begin{center}
    \includegraphics[angle=0,width=\textwidth]{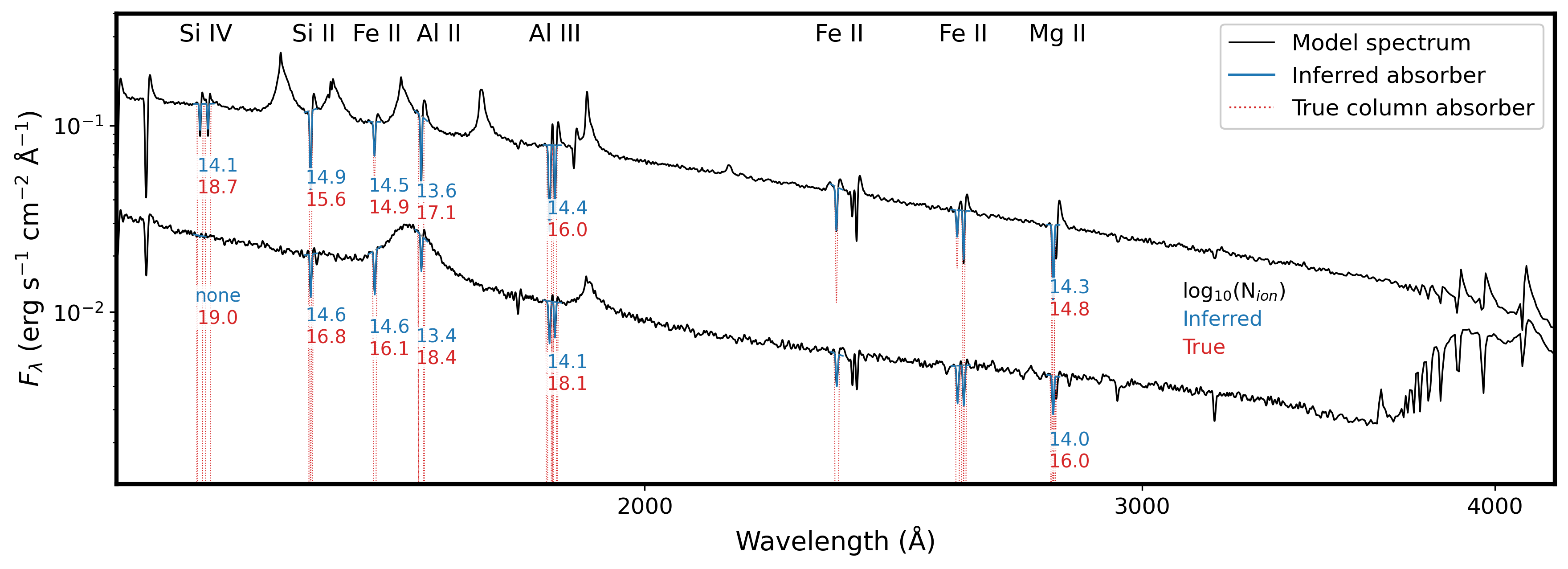}
\end{center}
\vspace{-0.6cm}
\caption{\sirocco model spectrum for lowest and highest column model in Fig.~\ref{fig:al}. The blue curve shows the absorber implied by the apparent column, and the red dotted curve the same absorber carrying the model's true radial column. Numbers give $\log(N_{\rm ion}/{\rm cm^{-2}})$ of each ion, inferred (blue) and true (red). At low column \ion{Mg}{ii} and \ion{Fe}{ii} are recovered to within 0.5\,dex, whereas \ion{Al}{ii}, \ion{Al}{iii} and \ion{Si}{iv} are saturated. At high column every line is saturated and the inferred columns fall 1-6\,dex below the true ones. }
\label{fig:al_spec}
\end{figure*}

\section{Sensitivity to black hole masses in the \sirocco models}\label{app:bh}

The fiducial models are defined by a $L=10^{44}$\,erg\,s$^{-1}$ blackbody with $3\times10^{4}$\,K irradiating a cocoon initialised at a radius $10^{16}\,{\rm cm}$ and $\tau_{\rm es}$ of a few. Fig.~\ref{fig:UV} employs a spherical cocoon with gas expanding radially at 100--1000\,km\,s$^{-1}$ \citep[see][]{Sneppen2026d} and Figs.~\ref{fig:abs_centroid}, \ref{fig:abs_centroidII} and \ref{fig:NH} 
use an axisymmetric cocoon with a ballistic outflow decelerating from 600 to 90\,km\,s$^{-1}$. Currently, a formal parameter search is computationally prohibitive, so we use these examples to illustrate the generic UV constraints (e.g.\ absorber centroids for velocity-structures) or to elucidate atomic data arguments (e.g.\ emission strengths for N/C). In earlier models, we were agnostic to the velocity-law, as choosing between stellar, homologous or marginally unbound winds did not strongly affect the emergent spectral continuum and the H and He absorbers only probed the characteristic line-of-sight velocity \citep{Sneppen2026,Kaaz2026}. Parameter exploration is detailed in \citet{Sneppen2026,Sneppen2026d} and the presented specific model spectra are typical of this part of the simulation grid, rather than extreme cases. 

To gauge absorber kinematics constraints on the black hole mass, we rescaled all radii of the fiducial \sirocco model by a factor $s=0.1{-}10$ at fixed luminosity, lowering the density inside the clumps by $s^{-2}$ (to keep the ionisation similar) and raising their volume filling factor by $s$ (to keep the column density fixed). The models thereby still obey the same observational constraints, keeping the absorber ladder preserved and reproducing a similar UV spectrum. The effective masses $M_{\rm BH}(1-\Gamma)$ differ by two orders of magnitude, but the inferred black hole mass nevertheless varies only by a factor of few (see Fig.~\ref{fig:BH_mass}). That is because i) the deceleration measures only the difference between gravity and radiation force,  $M_{\rm BH} (1-\Gamma)$ and ii) the luminosity constrains $\Gamma M_{\rm BH}$. The \sirocco inferred black hole mass is constrained by this intersection. Within the model framework, we are not strongly sensitive to the scale of the cocoon as radiation force and gravity both decrease proportionally with distance, while the exact geometrical structure does not affect the black hole gravitational potential. Within this model-space the dominant uncertainty becomes the intrinsic luminosity, which, with uncertainty at the factor of a few level, linearly propagates into the uncertainty in $M_{\rm BH}$. 
The mass thus relies on being able to infer the bolometric luminosity \citep[e.g.][]{Greene2025}. This is only strongly biased if dust dominates the reddening of the SED (instead of intrinsically red gas emission) and most of the energy is emitted through dust, in tension with most infrared observations \citep{PerezGonzalez2026}. 
Other interpretations of the deceleration could significantly bias the gravitational mass, so are also worth defining and testing against the data. For instance, a wind ploughing into dense ambient gas would slow in a momentum-conserving shell. Here the deceleration would be set by the ambient density rather than by gravity. It is not clear, however, how shocks would produce the ordered velocities of different ions, because the ambient medium would be both the reprocessing and the absorbing gas, with no clear velocity structure on its own.  

For Fig.~\ref{fig:BH_mass}, we include the comparison mass inferred from H$\alpha$ in G395M using local relations \citep{Greene2005} and using these relations when correcting for electron-scattering \citep{Rusakov2025}. The latter includes a broad Gaussian Doppler core, exponentially scattered wings and a Gaussian absorber, with the narrow lines set by the unresolved [O\,{\sc iii}]\,$\lambda5008$. The core broad FWHM is 300--400\,km\,s$^{-1}$, embedded within exponential wings which carry 70--80\% of the broad flux at $\tau_{\rm e}\approx1.1$--1.5. Using standard local relations \citep{Reines2015}, the core gives $\log(M_{\rm BH}/{\rm M}_\odot)=6.4\pm0.2$, versus 8.0 for a single broad Gaussian and 7.8 for two-Gaussian parametrization (both for a narrow line with free centroid). GN-2 is a fairly bright LRD with a H$\alpha$ luminosity $\times7$ and $\times10$ brighter than the median source in \citet{Sneppen2026} and \citet{deGraaff2025b}, respectively. Thus, each estimator gives a value somewhat more massive compared to a typical LRD object.

\end{appendix}

\end{document}